\documentclass[amsfonts,amsmath,prd,preprint,nofootinbib]{revtex4}%[aps,amsfonts,amsmath,prd,preprint,nofootinbib]{revtex4}

\newcommand{\beq}{\begin{equation}}
\newcommand{\eeq}{\end{equation}}

\usepackage{epsfig,bbm,cancel,ulem}
\usepackage{xcolor}
\usepackage[breaklinks=true]{hyperref}
\usepackage{graphicx}
\usepackage{amsmath}
\usepackage{amssymb}

\usepackage{wrapfig}
\usepackage{multirow}%
\usepackage{wasysym}
\usepackage[mathscr]{eucal}
\usepackage{amsfonts}%
\usepackage{amsthm}%
\usepackage{mathrsfs}%
\usepackage[title]{appendix}%
\usepackage{xcolor}%
\usepackage{textcomp}%
\usepackage{booktabs}%
\usepackage{algorithm}%
\usepackage{algorithmicx}%
\usepackage{algpseudocode}%
\usepackage{listings}%
\usepackage{subcaption}
\usepackage{tikzpagenodes}
\usepackage{tikz}
\usetikzlibrary{calc}
\usepackage{mwe}
\usepackage{setspace}
\usepackage{enumitem}
\usepackage{tabularray}
\usepackage{tablefootnote}
\usepackage{float}
\begin{document}

\title{Gravitational field and lensing of a circular chiral vorton}%{Weak-field Gravity and Lensing of Circular Chiral Vorton}
\author{Leonardus B.~Putra}
\email{leonardus.brahmantyo@ui.ac.id}
\author{H.~S.~Ramadhan\footnote{Corresponding author.}}
\email{hramad@sci.ui.ac.id}
\affiliation{Departemen Fisika, FMIPA, Universitas Indonesia, Depok, 16424, Indonesia.}
\def\changenote#1{\footnote{\bf #1}}

\begin{abstract}
We derive the metric of a circular chiral vorton in the weak field limit. The object is self-supporting by means of its chiral current. A conical singularity with deficit angle, identical to that of straight string with the same linear mass density, is present at the vorton's core. We find that the metric is akin to the electromagnetic $4$-potential of a circular current wire loop, illustrating the concept of gravito-electromagnetism. Surprisingly we find that the solution asymptotically mimics a Kerr-like naked singularity with mass $M_v=4\pi R\mu$ and spin parameter $a=R/2$. Finally, we also simulate the gravitational lensing images by solving the corresponding null geodesic equations. This reveals interesting properties of the images, such as the simultaneous creation of a minimally distorted source image and its Einstein ring, as well as the formation of double images on the back side of the ring.% such as the simultaneous creation of minimally distorted source image and its Einstein ring, as well as the creation of double images of the back side of the ring.

\end{abstract}

\maketitle
\thispagestyle{empty}
%\section{Introduction}
\setcounter{page}{1}

\section{Introduction}
\label{sec1}

Kibble mechanism suggests that cosmic strings could have formed in the early universe, most likely during the {\it Grand Unified} phase transition, when the corresponding vacuum manifold is non simply-connected, $\pi_1\left(\mathcal{M}\right)\cong\mathbb{Z}$~\cite{Kibble:1976sj, Vilenkin:1984ib, CSBookVilenkin:2000jqa}. Topology dictates that, upon formation, strings must either extend infinitely or form closed loops. In both cases, assuming the string's curvature is significantly larger than its thickness, its dynamics can then be effectively approximated by the Nambu-Goto action~\cite{Nambu, Goto:1971ce}.

Witten was the first to suggest that strings could be superconducting~\cite{Witten:1984eb}, 
 which he showed specifically within a $U(1)\times U(1)$ gauged theory. The dynamics of superconducting strings in general can be very complicated~\cite{Copeland:1987th}, but Carter and Peter showed that in the chiral limit, $j^{a}j_{a}=0$, the corresponding equations of motion can be exactly solved~\cite{Carter:1999hx} (see also~\cite{Blanco-Pillado_PhysRevD.63.103513, Davis_PhysRevD.62.083516}). In particular, the current flowing through the string generates angular momentum that, when sufficiently large, can stabilize the loop against collapse due to its tension. This spinning soliton-like object is known as a vorton~\cite{Davis:1988jq, DAVISShellardVorton1989209}. Extensive studies have been conducted on their formation~\cite{Martins:1998gb, Battye:2008mm, Auclair:2020wse} and stability~\cite{Carter:1990sm, Lemperiere:2003yt, Carter:1993wu, Battye:2021sji, Battye:2021kbd}.
 
Once hypothesized to seed galaxy formation long ago~\cite{Vilenkin:1981iu}, cosmic strings have recently regained interest due to the discovery of gravitational waves~\cite{LIGOScientific:2013tfe, Belahcene:2019ric}. This stems from predictions that oscillating loops of cosmic strings could be efficient sources of gravitational radiation~\cite{Vachaspati:1984gt, Vachaspati:1987sq}. The gravitational field for a circular loop was studied by Frolov, Israel, and Unruh (FIU)~\cite{Frolov:1989er}. They show that the metrics suffer from deficit angle $\delta\varphi=8\pi G\mu$, with $\mu$ is the string tension. The enormous tension forces a loop to oscillate, while the FIU metric, on the other hand, is stationary. Some authors postulate the need for an external pressure~\cite{Hughes_PhysRevD.47.468} or a revolving mechanism~\cite{McManus_PhysRevD.47.1491} to prevent the loop from collapsing.

In our view, the formation of a vorton, which does not require any external mechanism, is the most natural stabilization scenario for a stationary string loop. Surprisingly, there has been very little study on the gravitational field of vortons. To the best of our knowledge, the only study on the gravitational field of vortons to date is by Kunz, Radu, and Subagyo~\cite{KunzRaduBintoro_PhysRevD.87.104022}. They obtained a gravitating $U(1) \times U(1)$ global vorton by numerically solving the full Einstein and scalar field equations. For certain parameter values, the vorton has an ergoregion surrounding it, indicating instability~\cite{Cardoso:2007az}. Asymptotically, the metric approaches that of a Kerr black hole. 

The purpose of this work is to further elaborate on the gravitational field of vortons and its astrophysical impact. In this paper, we study the spacetime around a circular chiral vorton in the weak-gravitational field limit and its properties. These linearized solutions will then be used to calculate its gravitational lensing effects. The paper is organized as follows. In Sec.~\ref{sec:gravvorton}, we derive the metric of a circular chiral vorton using the weak-field and thin-string approximations. We begin by discussing the appropriate metric ansatz for a rotating loop, the energy-momentum tensor for the chiral vorton in the thin-string approximation, and finally the weak-field solution including its asymptotic behavior. In Sec.~\ref{sec:prop}, we explore the physical properties of the metric solution we obtain. Sec.~\ref{sec:gravlens} addresses the lensing signatures of the vorton. We numerically solve the corresponding null geodesic equations to determine the light-bending patterns caused by a circular vorton with various plane orientations relative to the optical axis. Finally, we summarize our findings in Sec.~\ref{sec:conc}.

\section{Gravitational Field of Vorton}
\label{sec:gravvorton}

\subsection{The metric ansatz}
%\label{sec:metricansatz}

In general, a cosmic string loop cannot be stationary due to its tension. However, its collapse and re-expansion can be prevented by the right amount of angular momentum along the string. Therefore, the most general metric for a rotating string loop is axially-symmetric with two commuting Killing vectors $\partial_t$ and $\partial_\phi$. In cylindrical coordinate  it reads~\cite{McManus_PhysRevD.47.1491, Hughes_PhysRevD.47.468}
\begin{equation}
\label{vortoncylansatz}
    ds^2=-e^{2\nu(r,z)}dt^2+e^{2\zeta(r,z)-2\nu(r,z)}r^2\left(d\phi-A(r,z)dt\right)^2+e^{2\eta(r,z)-2\nu(r,z)}\left(dr^2+dz^2\right).
\end{equation}
Assuming that all the metric functions ($\nu$, $\zeta$, $\eta$, and $A$) are weak field of order $G\mu$, the Einstein tensor components are%in first order is 
 ~\cite{Hughes_PhysRevD.47.468}
\begin{subequations}
\begin{eqnarray}
\label{Gtt}
G_t^t&=&-2\nabla^2\nu+\nabla^2\zeta+\frac{1}{r}\zeta_r+\Tilde{\nabla}^2\eta,\\ %\end{equation}
%\begin{equation}
    G_\phi^t&=&-\frac{1}{2}r^2\left(\nabla^2A+\frac{2}{r}A_r\right),\\
%\end{equation}
%\begin{equation}
G_\phi^\phi&=&\Tilde{\nabla}^2\eta,\\ %\end{equation}
%\begin{equation}
    G_r^r&=&\zeta_{zz}+\frac{1}{r}\eta_r,\\
%\end{equation}
%\begin{equation}
    G_r^z&=&-\zeta_{zr}-\frac{1}{r}\left(\zeta_z-\eta_z\right),\\
%\end{equation}
%\begin{equation}
\label{Gzz}
    G_z^z&=&\zeta_rr-\frac{1}{r}\left(\eta_r-2\zeta_r\right).
\end{eqnarray}
\end{subequations}
Here, $\nabla^2\equiv\partial_r^2+\frac{1}{r}\partial_r+\partial_z^2$ is the Laplacian in cylindrical coordinate, while $\Tilde{\nabla}^2\equiv\partial_r^2+\partial_z^2$ is the Laplacian in Cartesian coordinate with $(r,z)$ in place of $(y,z)$. 

In the toroidal coordinates $(t,\phi,\sigma,\psi)$, using transformations\footnote{Brief description of toroidal coordinate system is given in Appendix~\ref{Tor}.} 
\begin{eqnarray}
\label{transztor}
z&=&\frac{R}{N(\sigma,\psi)^2}\sin{\psi},\\
%\end{equation}
%\begin{equation}
\label{transrtor}
r&=&\frac{R}{N(\sigma,\psi)^2}\sinh{\sigma},
\end{eqnarray}
with
\begin{equation}
    N(\sigma,\psi)^2\equiv \cosh{\sigma}-\cos{\psi},
\end{equation}
the metric becomes
\begin{equation}
\label{vortontoransatz}
    ds^2=-e^{2\nu}dt^2+R^2N^{-4}\sinh^2{\sigma}e^{2\zeta-2\nu}\left(d\phi-Adt\right)^2+R^2N^{-4}e^{2\eta-2\nu}\left(d\sigma^2+d\psi^2\right).
\end{equation}
The weak field assumption of the ansatz means that the solutions of function $\nu$, $A$, $\eta$, and $\zeta$, are of order of $O(G\mu)$.

\subsection{The energy momentum tensor}

In the thin-string approximation, the string's dynamics is well-described by the Nambu-Goto equations
\begin{equation}
\ddot{x}^{\mu}-x^{\mu\prime\prime}=0,
\end{equation}
The corresponding energy momentum tensor is~\cite{Blanco-Pillado_PhysRevD.63.103513,Davis_PhysRevD.62.083516}
\begin{equation}
\label{emtensorchiral}
T^{\mu\nu}=\mu\int d\eta \left(\Dot{x}^\mu\Dot{x}^\nu-x'^\mu x'^\nu\right)\delta^{(3)}\left(\Vec{y}-\Vec{x}(t,\eta)\right).
\end{equation}

The solutions for the chiral string loop are of the form
\begin{equation}
    x^\mu=\left(t,\Vec{x}\right),
\end{equation}
where
\begin{equation}
    \Vec{x}=\Vec{a}(q)+\Vec{b}(\eta).
\end{equation}
Here $\Vec{a}$ and $\Vec{b}$ are two 3-vectors satisfying the constraints
\begin{equation}
    %\begin{split}
        |\Vec{a}|^2=1,\ \ \ \
        |\Vec{b}|^2=k^2,
   % \end{split}
\end{equation}
where $k$ is some constant, with $q=t+\sigma$ and $\eta=t-\sigma$. Vorton is the class of stationary string loop solutions satisfying $k=0$. 

A circular vorton solution in cylindrical coordinate $(t,\phi,r,z)$ can be written as (similar to the axial coordinate in~\cite{CCSLKerrPhysRevD.79.065029})
\begin{equation}
\label{gravcircvortsol}
    x^\mu=\bigg(t,\ \frac{q}{2R}+f(t),\ R,\ 0\bigg),
\end{equation}
which describes a vorton of constant radius $R$ at $z=0$, with its center at the origin. The axial coordinate is the only one that depends on $\sigma$, and the relation is linear to account for the axial symmetry of the system. Here the function $f(t)$ is added to take into account the constraints
\begin{equation}
    \Dot{x}^\mu x'_\mu=\frac{1}{4}(1-k^2),
\end{equation}
where we take $k=0$ for the case of vorton
\begin{equation}
\label{vortconst}
    \Dot{x}^\mu x'_\mu=\frac{1}{4}.
\end{equation}
Substituting the solution \eqref{gravcircvortsol} to the constraint \eqref{vortconst}, and using the weak field limit assumption of the ansatz \eqref{vortoncylansatz}, we have
\begin{equation}
    \Dot{f}(t)=\mathcal{O}(G\mu).
\end{equation}
And therefore the (time dependent) shift of the spacelike string axial coordinate is of order $G\mu$. 

Ignoring the higher order term of $G\mu$, the nonzero mixed energy-momentum tensor components in this coordinate would be
\begin{equation}
\label{16}
    GT_t^t=-2G\mu\delta(r-R)\delta(z),
\end{equation}
and
\begin{equation}
\label{17}
    GT^t_\phi=G\mu R\delta(r-R)\delta(z).
\end{equation}
Note how in our case here, the tension of the string approaches zero, so are all the other $ij$ components of the stress-energy tensor, just like the case of the infinite straight chiral string \cite{GravityStraightChiralStringSteer:2000jn}.

\subsection{The weak-field solution}

%The metric solution can be obtained using the Einstein Field Equation
%\begin{equation}
%    G^\mu_\nu=8\pi GT^\mu_\nu.
%\end{equation}
%Using the Einstein tensor and the energy-momentum tensor, we get the field equations
The Einstein's equations read
\begin{eqnarray}
\label{EGtt}
-2\nabla^2\nu+\nabla^2\zeta+\frac{1}{r}\zeta_r+\Tilde{\nabla}^2\eta&=&-16\pi G\mu\delta(r-R)\delta(z),\\
%\end{equation}
%\begin{equation}
\label{EGtphi}
-\frac{1}{2}r^2\left(\nabla^2A+\frac{2}{r}A_r\right)&=&8\pi G\mu R\delta(r-R)\delta(z),\\
%\end{equation}
%\begin{equation}
\label{EGphiphi}
\Tilde{\nabla}^2\eta&=&0,\\
%\end{eqnarray}
%begin{equation}
\label{EGrr}
    \zeta_{zz}+\frac{1}{r}\eta_r&=&0,\\
%\end{equation}
%\begin{equation}
\label{EGzr}
    -\zeta_{zr}-\frac{1}{r}\left(\zeta_z-\eta_z\right)&=&0,\\
%\end{equation}
%\begin{equation}
\label{EGzz}
    \zeta_{rr}-\frac{1}{r}\left(\eta_r-2\zeta_r\right)&=&0.
\end{eqnarray}

We can convince ourselves that the Eqs.~\eqref{EGtt}-\eqref{EGzz} are consistent. Eqs.~\eqref{EGzr} and \eqref{EGzz} gives
\begin{equation}
\label{etaz}
    \eta_z=\zeta_z+r\zeta_{zr},
\end{equation}
and
\begin{equation}
\label{etar}
    \eta_r=r\zeta_{rr}+2\zeta_r,
\end{equation}
respectively. Differentiating Eq.~\eqref{EGrr} and~\eqref{etaz} wrt $r$ gives
%\begin{equation}
%    \eta_r+r\zeta_{zz}=0,
%\end{equation}
%\textcolor{blue}{which, under differentiation wrt $r$, gives}
\begin{eqnarray}
%\begin{split}
%\eta_{rr}+\zeta_{zz}+r\zeta_{zzr}&=0\\
\eta_{rr}+\partial_z\left(\zeta_z+r\zeta_{zr}\right)&=&0,\\
\eta_{zz}-\partial_z\left(\zeta_z+r\zeta_{zr}\right)&=&0,
%\end{split}    
\end{eqnarray}
respectively. This immediately leads to
\begin{equation}
\eta_{rr}+\eta_{zz}=\Tilde{\nabla}^2\eta=0,
\end{equation}
which is consistent with \eqref{EGphiphi}.
%\par 

Integrating Eq.~\eqref{etar} yields
%\begin{equation}
%    \int\eta_r dr=\int r\zeta_{rr}dr+2\int \zeta_r dr,
%\end{equation}
\begin{equation}
\label{etaf}
    \eta=r\zeta_r+\zeta+f(z),
\end{equation}
while integrating Eq.~\eqref{etaz} gives%to get
%\begin{equation}
%    \int\eta_z dz=\int\zeta_z dz+r\int\zeta_{rz}dz
%\end{equation}
\begin{equation}
\label{etag}
    \eta=r\zeta_r+\zeta+g(r).
\end{equation}
%From Eqs.~\eqref{etaf} and \eqref{etag}, 
Thus, %we can conclude that
\begin{equation}
\label{eta0}
    \eta=r\zeta_r+\zeta+\eta_0,
\end{equation}
with $\eta_0$ a constant. From~\eqref{EGphiphi}, the ansatz can be written in the form
\begin{equation}
    \eta=X(r)Z(z),
\end{equation}
whose solution can only either be a linear or a combination of harmonic and hyperbolic functions. Either way, it violates the asymptotic flatness. Therefore, the function must be constant, which can be absorbed into $\eta_0$,
\begin{equation}
\label{etasemifinal}
    \eta=\eta_0.
\end{equation}
Hence, with \eqref{eta0} we have
\begin{equation}
\eta_0%=r\zeta_r+\zeta
=\partial_r\left(r\zeta\right)\ \ \ \rightarrow\ \ \ \eta_0 r = \zeta r - h(z)
\end{equation}
%\begin{equation}
%    \eta_0 r = \zeta r - h(z),
%\end{equation}
and thus
\begin{equation}
    \zeta = \eta_0 +\frac{h(z)}{r}.
\end{equation}
Further, regularity at the core demands $h(z)=0$, thus
\begin{equation}
\label{zetasemifinal}
    \zeta = \eta_0.
\end{equation}
This is also evident from \eqref{EGrr}, where we find $\zeta_{zz}=0$. Consequently, $h(z)$ must be a constant to satisfy the asymptotic flatness condition.

Using Eqs.~\eqref{etasemifinal} and \eqref{zetasemifinal}, Eq.~\eqref{EGtt} becomes
\begin{equation}
    \nabla^2\nu=8\pi G\mu\delta(r-R)\delta(z).
\end{equation}
This is the Poisson's equation for the electrostatic potential of a ring charge. Following \cite{Hughes_PhysRevD.47.468}, the solution in the toroidal coordinate is
\begin{equation}
\label{nufinaltor}
\nu(\sigma,\psi)=-4\sqrt{2}G\mu N(\sigma,\psi)\frac{K\left(\tanh{\frac{\sigma}{2}}\right)}{\cosh{\frac{\sigma}{2}}},
\end{equation}
where
\begin{eqnarray}
K\left(a\right)\equiv\int_0^{\frac{\pi}{2}}\frac{d\theta}{\sqrt{1-a\sin^2{\theta}}}
\end{eqnarray}
is the {\it complete elliptic integral of the first kind}. In cylindrical coordinate it takes the form
\begin{equation}
\label{nufinalcyl}
    \nu(\sigma,\psi)=-4 \sqrt{2} G\mu \sqrt{\frac{R}{r} \sinh \left(\frac{1}{2} \ln \left(\frac{(r+R)^2+z^2}{(r-R)^2+z^2}\right)\right)} \frac{ K\left(\tanh \left(\frac{1}{4} \ln \left(\frac{(r+R)^2+z^2}{(r-R)^2+z^2}\right)\right)\right)}{\cosh \left(\frac{1}{4} \ln \left(\frac{(r+R)^2+z^2}{(r-R)^2+z^2}\right)\right)}.
\end{equation}
The {\it complete elliptic integral} $K(a)$ exhibits a singularity at $a=1$, which corresponds to $\sigma=\infty$, the coordinate of the vorton's core. As $\sigma$ increases, $K(a)$ decreases, indicating that the field diverges at the core and diminishes with distance from it. This divergence is expected to disappear once the thin-vorton approximation is relaxed. Furthermore, the inclusion of higher-order corrections of $G\mu$ does not affect the first-order term, and we do not anticipate significant contributions from higher-order terms, such as $O(G^2\mu^2)$ and above.

\begin{figure}
\centering
\includegraphics[width=0.70\textwidth]
{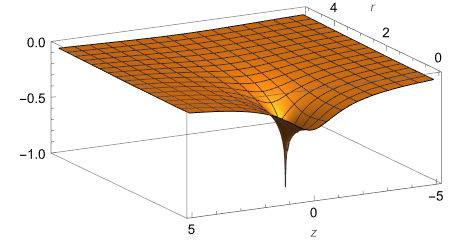}
\caption{3D Plot of $\nu(r,z)$ for $G\mu=0.03$.}
\label{fig:nu}
\end{figure}

The function $A(r,z)$ is not necessarily needed for calculating the deficit angle, however we could also solve it in order to have the full spacetime profile of circular vorton. Eq.~\eqref{EGtphi} can be re-expressed into a nicer form
\begin{equation}
    \nabla^2A+\frac{2}{r}A_r=-16\pi \frac{G\mu}{R}\delta(r-R)\delta(z).
\end{equation}
Defining $W\equiv rA$, it becomes
\begin{equation}
\label{weq}
    \nabla^2W-\frac{1}{r^2}W=-16\pi G\mu\delta(r-R)\delta(z).
\end{equation}
One can easily recognize that the equation above just resembles the $\varphi$-component of the Ampere equation for the magnetic vector potential with a current source $\Vec{J}=I\delta(r-R)\delta(z)\hat{\varphi}$.
%It is easy to see that the function $W(r,z)$ is just the $\varphi$-component of the magnetic vector potential of a circular ring current of radius $R$, where the equation $\nabla\times\Vec{B}=\mu_0 \Vec{J}$ with $\Vec{J}=I\delta(r-R)\delta(z)\hat{\varphi}$ and $\Vec{B}=\nabla\times\Vec{A}$ yields the same form of equation as Eq.~\eqref{weq}, only with $A_\phi$ in place of $W$.% 
Following Jackson~\cite{Jackson:1998nia}, we have
\begin{eqnarray}
\label{Afinalanalytic}
%    \begin{split}
    A(r,z)&=&\frac{16G\mu R}{r\sqrt{z^2+(r+R)^2}}\bigg[\left(\frac{z^2+(r+R)^2}{2rR}-1\right)K\left(\frac{4rR}{z^2+(r+R)^2}\right)\nonumber\\%\right.\\
    %&\left.
    &&-\frac{z^2+(r+R)^2}{2rR}E\left(\frac{4rR}{z^2+(r+R)^2}\right)\bigg],   
%\end{split}
\end{eqnarray}
or, in the toroidal coordinate 
\begin{equation}
A\left(\sigma,\psi\right)=\frac{8\sqrt2G\mu}{R}\frac{N^3\left(\sigma,\psi\right)}{\sinh^2{\sigma}}\bigg[e^{-\frac{\sigma}{2}}\cosh{\sigma}K\left(2e^{-\sigma}\sinh{\sigma}\right)-e^\frac{\sigma}{2}E\left(2e^{-\sigma}\sinh{\sigma}\right)\bigg],
\end{equation}
with
\begin{eqnarray}
E\left(k^2\right)\equiv\int_0^{\frac{\pi}{2}}\sqrt{1-k^2\sin^2{\theta}}d\theta,
\end{eqnarray}
the {\it complete elliptic integral of the second kind}. This solution can also be obtained by solving Eq.~\eqref{weq} directly using the Green's function method, as shown in the Appendix~\ref{DerA}.

\begin{figure}
	\centering
\includegraphics[width=0.70\textwidth]
{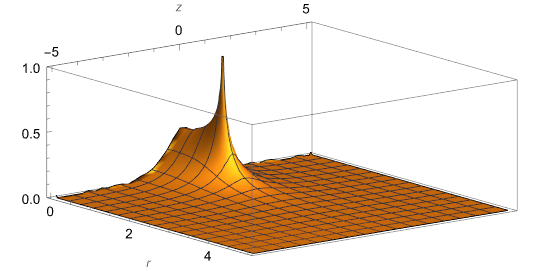}
	\caption{3D Plot of $A(r,z)$ for $G\mu=0.03$.}
	\label{fig:A}
\end{figure}

\begin{figure}
     \centering
     \begin{subfigure}[b]{0.45\textwidth}
         \centering
\includegraphics[width=\textwidth]{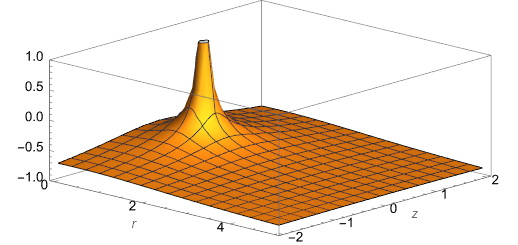}
         \caption{$g_{tt}(r,z)$}
         \label{fig:gtt}
     \end{subfigure}
     \hfill
     \begin{subfigure}[b]{0.45\textwidth}
         \centering
\includegraphics[width=\textwidth]{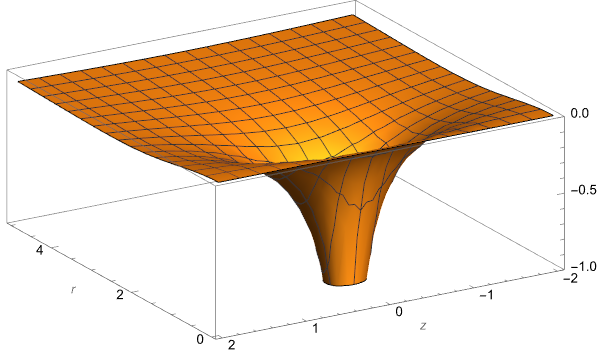}
\caption{$g_{t\phi}(r,z)$}
         \label{fig:gtphi}
     \end{subfigure}
     \hfill
     \begin{subfigure}[b]{0.45\textwidth}
         \centering
\includegraphics[width=\textwidth]{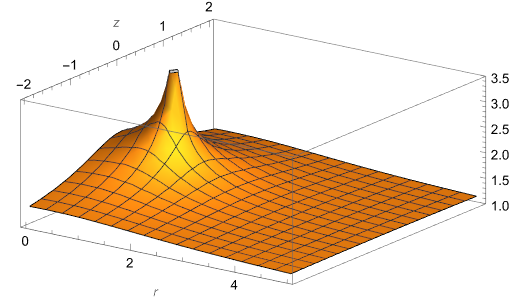}
         \caption{$g_{rr}(r,z)=g_{zz}(r,z)$}
         \label{fig:grr}
     \end{subfigure}
     \hfill
     \begin{subfigure}[b]{0.45\textwidth}
         \centering
\includegraphics[width=\textwidth]{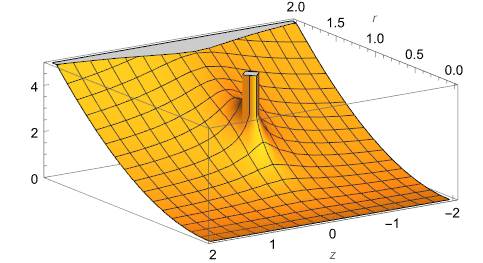}
\caption{$g_{\phi\phi}$}
        \label{fig:gphiphi}
     \end{subfigure}
        \caption{Complete metric of the circular vorton for $G\mu=0.03$ and $R=1$.}
        \label{fig:vortonmetric}
\end{figure}

In Figs.~\ref{fig:nu}-\ref{fig:vortonmetric} we show the profiles of the metric functions. It can be seen that % from \ref{fig:vortonmetric}, \ref{fig:A}, and \ref{fig:nu} that 
all functions % (including $\nu$ and $A$)  
diverge at $(r,z)=(1,0)$, %that is 
precisely at the string cross section of $R=1$. This is expected, as it is the natural behavior of a field with delta function distribution% (which will have some Green's functions as solutions), 
coming from the thin string approximation. It does indicate the presence of ring singularity (or ringularity) without a horizon. The ringularity is expected to be smoothed out with the order of the string thickness $\delta$ once we consider the full gravitating thick vorton (like in the case of \cite{KunzRaduBintoro_PhysRevD.87.104022}).%, we could expect this singularity to be smoothed out with the order of the string thickness $\delta$.

\subsection{Asymptotic behaviour of the metric}

The asymptotic expansion of $\nu$ near the string ($\sigma\rightarrow\infty$ in the toroidal coordinate) can be obtained using the asymptotic behavior of the hyperbolic functions and elliptic integral,
\begin{equation}
\cosh{\frac{\sigma}{2}}\rightarrow\frac{1}{2}e^{\sigma/2},\ \ \ N(\sigma,\psi)\rightarrow\sqrt{\frac{1}{2}}e^{\sigma/2},
\end{equation}
%\begin{equation}
%    \cosh{\frac{\sigma}{2}}\rightarrow\frac{1}{2}e^{\sigma/2},
%\end{equation}
and
\begin{equation}
    K\left(\tanh{\frac{\sigma}{2}}\right)\rightarrow\frac{\sigma}{2}.
\end{equation}
Thus,
\begin{equation}
\label{nuinf}
\nu(\sigma\rightarrow\infty)=-4G\mu\sigma.
\end{equation}
\par Meanwhile, the asymptotic behavior far from the vorton would be
\begin{equation}
\label{farasymptoticnu}
\nu\left(\sqrt{r^2+z^2}\rightarrow\infty\right)\rightarrow-\frac{4\pi G\mu R}{\sqrt{r^2+z^2}}.
\end{equation}
We can therefore deduce from the ansatz metric \eqref{vortoncylansatz} that at infinity, we get
\begin{equation}
    \eta_0\to0,
\end{equation}
%We then get the asymptotic form of $g_{rr}$, which is
which implies
\begin{equation}
g_{rr}\rightarrow\left(1-\frac{8\pi G\mu R}{\sqrt{r^2+z^2}}\right)^{-1}.
\end{equation}
Defining $\rho=\sqrt{r^2+z^2}$ and expressing the vorton mass as
\begin{equation}
\label{vortonmass}
M_v\equiv4\pi R\mu,
\end{equation}
we get
\begin{equation}
\label{grrschwarszhcild}
    g_{rr}\rightarrow\left(1-\frac{2GM_v}{\rho}\right)^{-1},
\end{equation}
which is just the Schwarzschild metric of a point mass $M_v$, with $M_v$ the mass of a ring of linear density $\mu$ and radius $2R$. This is expected from the formulation of chiral vorton, where the current energy is equal to the mass term $\mu L$. Hence the total rest energy is twice the mass term $E_v=2\mu L$, where $L$ is the invariant length of the vorton, which in the circular case would be $2\pi R$.

The asymptotic expansion of $A$ near the string $(\sigma \rightarrow\infty)$ can be found by taking
\begin{equation}
    K\left(1-e^{-2\sigma}\right)\rightarrow \sigma,\ \ \ E\left(1-e^{-2\sigma}\right)\rightarrow 1,
\end{equation}
%\begin{equation}
%    E\left(1-e^{-2\sigma}\right)\rightarrow 1,
%\end{equation}
%\begin{equation}
%\sinh{\sigma}\rightarrow\frac{e^\sigma}{2},\ \ \ \cosh{\sigma}\rightarrow\frac{e^\sigma}{2},
%\end{equation}
%\begin{equation}
%\cosh{\sigma}\rightarrow\frac{e^\sigma}{2},
%\end{equation}
which gives
\begin{equation}
A\left(\sigma\rightarrow\infty,\psi\right)\to\frac{8G\mu\sigma}{R}.
\end{equation}
This confirms the behavior that $A(r,z)$ diverges at the vorton core.

The asymptotic expansion at infinity of $A$ is
\begin{equation}
A(\rho\rightarrow\infty)\to\frac{4\pi G\mu R^2}{\rho^3}+O\left(\frac{1}{\rho^4}\right),
\end{equation}
where $\rho=\sqrt{r^2+z^2}$ is the radial position in spherical coordinate. Thus, the expansion of $g_{t\phi}$ at infinity in spherical coordinate is
\begin{equation}
    g_{t\phi}(\rho\rightarrow\infty)\to-\frac{4\pi G\mu R^2}{\rho}\sin^2{\theta}+O\left(\frac{1}{\rho^2}\right),
\end{equation}
which coincides with the expansion of Kerr metric at infinity
\begin{equation}
    g_{t\phi}(\rho\rightarrow\infty)\to-\frac{2GJ}{\rho}\sin^2{\theta}+O\left(\frac{1}{\rho^2}\right),
\end{equation}
with
\begin{equation}
\label{kerrJ}
    J=aM,
\end{equation}
the angular momentum of the black hole of spin parameter $a$ and mass $M$. The vorton's angular momentum can then be identified as
\begin{equation}
\label{vortonangularmomentum}
    J_v=\frac{1}{2}RM_v,
\end{equation}
which is similar to Kerr~\eqref{kerrJ}, only with
\begin{equation}
    \label{vortonspinvalue}
    a=\frac{R}{2}.
\end{equation}
We can conclude that the circular vorton spacetime is asymptotically identical to the Kerr metric of a spinning black hole with mass $M_v=4\pi R\mu$ and spin parameter $a=R/2$. This raises an intriguing possibility: some observations of rotating black holes, particularly those involving nearby orbits, could potentially be attributed to vortons with the same mass. It is also worth noting that for the same amount of mass, the vorton would have a radius of:
\begin{eqnarray}
\label{vortonradiuswrtrs}
    R_v&=&\frac{GM_v}{4\pi c^2\alpha}=\frac{1}{8\pi\alpha}R_s,
%\end{equation}
%or
%\begin{equation}
\end{eqnarray}
where $R_s$ is the Schwarzschild radius of a black hole of the same mass, and $\alpha$ is the $G\mu$ value of the string. For a Grand Unified Theory (GUT) scale string, $\alpha$ is on the order of  $10^{-6}$. This implies that the vorton would have a radius $\sim10^{5}$ times that of a black hole of the same mass, which still falls within the far-field (Schwarzschild) approximation.

Interestingly, for a typical GUT-scale vorton, Eqs.~\eqref{vortonspinvalue}-\eqref{vortonradiuswrtrs} imply
\begin{equation}
\label{nakedsingularitycondition}
R_s^2 - 4a^2 < 0.
\end{equation}
Thus, the Kerr metric solution that the vorton asymptotically approaches corresponds to a naked singularity. A distant observer would find it challenging to distinguish between a vorton and a naked singularity. However, the spin of a black hole is typically measured using X-Ray spectroscopy of its accretion disk, as seen in \cite{SpinMeasurement10.1093/mnras/staa606}, rather than by the orbits of nearby stars. This is because spin measurement is a strong-field effect, while star orbits are a far-field effect~\cite{ObservingSpinReynolds:2019uxi,SpinConstraintsReynolds:2020jwt}. Furthermore, the proposed method for detecting a supposed naked singularity typically involves observing the object's strong field effects~\cite{NakedSingularityObsMethodPhysRevD.95.084024}. Unless a vorton could have an accretion disk—potentially formed from star matter orbiting too close to the vorton string core—it is not feasible to detect this apparent naked singularity, or more precisely, to distinguish it from a black hole of the same mass. Therefore, the discrepancy between the length scales at which the far-field approximation breaks down could help identify a vorton from nearby orbits. Since, like a black hole, a vorton produces no light on its own, this method could be crucial for distinguishing between the two objects. For example, the star S2 orbits the Sgr A* supermassive black hole with a closest approach of about $1400$ times the black hole's Schwarzschild radius \cite{S2SgrAGRAVITY:2018ofz}. This distance would not adhere to the far-field approximation if the object were a vorton. However, objects farther away from the vorton would. To accurately identify a vorton, the near vorton orbits must be studied more closely using the geodesic equations for massive objects. There could be several unique orbital solutions around a vorton, potentially including toroidal orbits, that could distinguish it from other dark sources. Further research in this area is needed to identify a vorton without relying solely on gravitational lensing, especially since resolving smaller scale vortons could be challenging.%In fact, there could be quite a few unique orbital solutions around a vorton (which may or may not include toroidal orbit) that could distinguish the object from other possibility of dark sources. Further research would be required in this area, which could be a way of identifying a vorton without having to rely on its gravitational lensing, since for the smaller scale vortons it could be harder to resolve.

\section{The Spacetime Properties}
\label{sec:prop}

\subsection{Frame dragging and ergoregion}

\begin{figure}
     \centering
     \begin{subfigure}[b]{0.45\textwidth}
         \centering
\includegraphics[width=\textwidth]{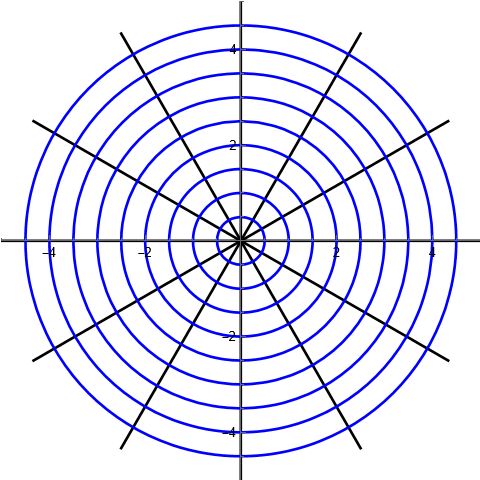}
         \caption{$t=0$}
     \end{subfigure}
     \hfill
     \begin{subfigure}[b]{0.45\textwidth}
         \centering
\includegraphics[width=\textwidth]{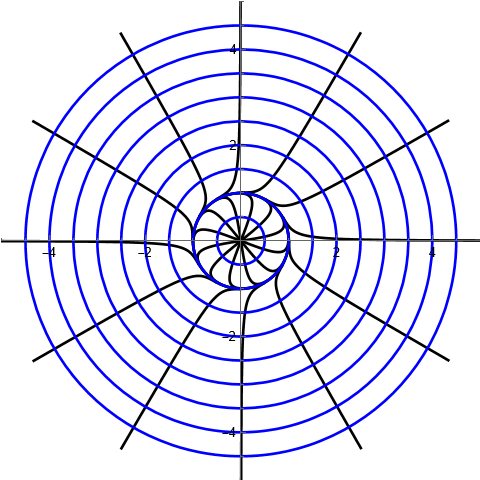}
         \caption{$t=1$}
     \end{subfigure}
     \hfill
     \begin{subfigure}[b]{0.45\textwidth}
         \centering
\includegraphics[width=\textwidth]{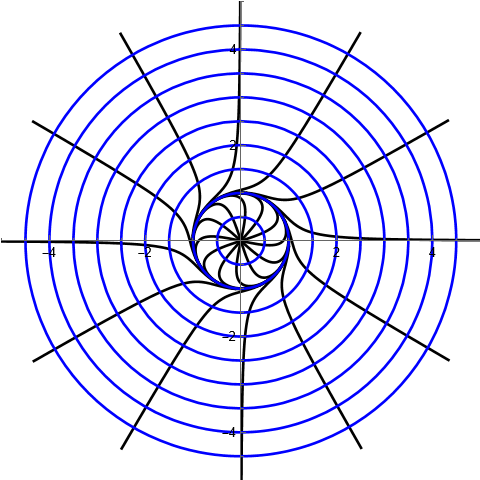}
         \caption{$t=2$}
     \end{subfigure}
     \hfill
     \begin{subfigure}[b]{0.45\textwidth}
         \centering
\includegraphics[width=\textwidth]{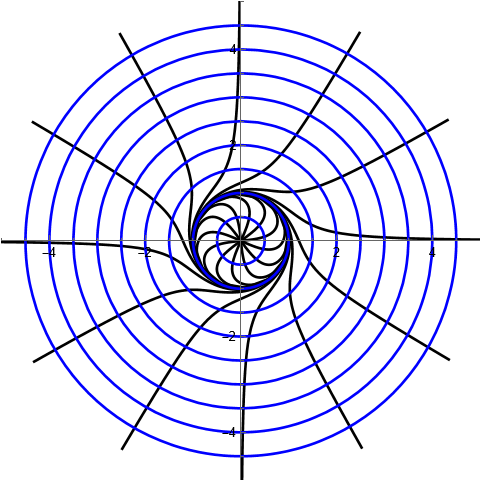}
         \caption{$t=4$}
     \end{subfigure}
        \caption{Polar grid of the spacetime with angular velocity $\Omega(r,z)$ at the vorton plane $z=0$ for $G\mu=0.2$ from $t=0$ to $t=4$.}
        \label{fig:framedragging}
\end{figure}

The metric solution represents a rotating spacetime, where an inertial frame of reference will rotate with angular velocity $\Omega(r,z)$ together with the spacetime. The angular frequency in the inertial frame is
\begin{equation}
    \Omega(r,z)=A(r,z),
\end{equation}
and thus  Eq.~\eqref{Afinalanalytic} dictates the angular velocity of the inertial frame at point $(r,z)$. A polar grid representation can be constructed by considering the parametric curve for several (equally spaced) value of $\phi$ as illustrated in Fig.~\ref{fig:framedragging}.
The figure represents the dragging of inertial frame around a circular vorton at $z=0$, which has the current running counter clockwise. As expected, the frame-dragging effect is more extreme near the string, and that the current drags the inertial frame in the same direction as the current. We also note that the analogue of this part of the metric and the magnetic vector potential of a circular current loop, together with the analogue of $\nu$ with the electric scalar potential of the same object is a manifestation of gravito-electromagnetism.

\begin{figure}
     \centering
     \begin{subfigure}[b]{0.45\textwidth}
         \centering
\includegraphics[width=\textwidth]{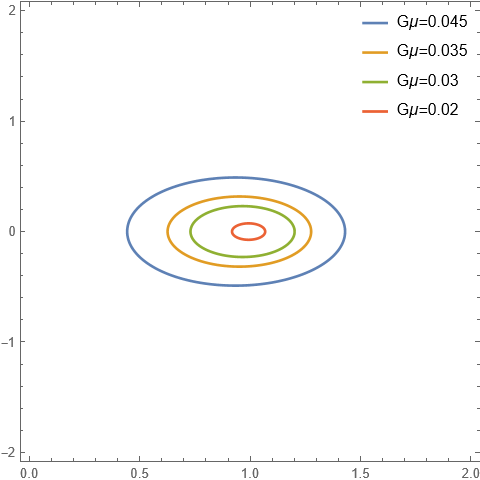}
\caption{}
    \label{fig:ergotoruscross}
     \end{subfigure}
     \hfill
     \begin{subfigure}[b]{0.45\textwidth}
         \centering
\includegraphics[width=\textwidth]{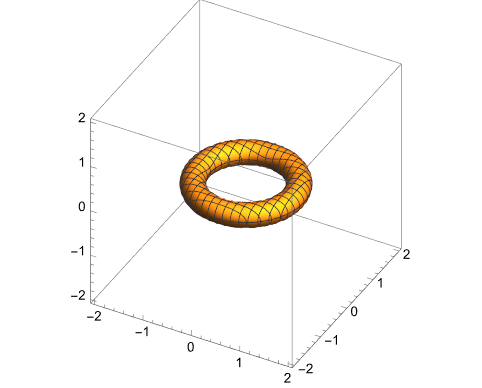}
\caption{}
         \label{fig:ergotorus}
     \end{subfigure}
     \caption{Curve of the surface of ergotorus cross section in $rz$-plane for various value of $G\mu$ (a), and 3D curve of ergotorus surface for $G\mu=0.03$ (b).}
    \label{fig:ergotoruspics}
\end{figure}

The ergoregion is the region between the event horizon and the ergosurface, where the event horizon is the surface $(r,z)$ of $g_{rr}=0$, and the ergosurface is the surface $(r,z)$ of $g_{tt}=0$. Since 
\begin{equation}
    g_{rr}=e^{-2\nu(r,z)}>0, \forall\ r,z,
\end{equation}
the solution contains no event horizon. However, the metric component $g_{tt}$
%\begin{equation}
%    g_{tt}=-e^{2\nu(r,z)}+e^{-2\nu(r,z)}r^2A^2(r,z),
%\end{equation}
does admit null solution
\begin{equation}
-e^{2\nu(r,z)}+e^{-2\nu(r,z)}r^2A^2(r,z)=0,
\end{equation}
since $g_{tt}<0$ for both $r\rightarrow\infty$ (condition~\eqref{farasymptoticnu}) and $r=0$, while $g_{tt}>0$ for both $r\to R$ and $z=0$. We can see this from the behavior of $\nu$ and $A$, where
\begin{eqnarray}
\nu\left(\sqrt{r^2+z^2}\rightarrow 0\right)&=&-4\pi G\mu,\\
%\end{equation}
%\begin{equation}
    r^2A(r\rightarrow0)&=&r W(r\rightarrow 0)=0,
\end{eqnarray}
and $A(r,z)\rightarrow+\infty$ for $(r\approx R,z=0)$. Since $g_{tt}$ changes sign from $(-+-)$ as $r$ increases, and the vorton cross-section lies within the region where $\text{sign}(g_{tt})=+1$, it is clear that the ergosurface must have a toroidal shape. It is called the {\it ergotorus}, illustrated in Fig~\ref{fig:ergotoruspics}.
The ergotorus is the region where the rotational energy of the vorton could potentially be extracted. %However, its presence suggests potential instability against scalar and electromagnetic perturbations, as discussed in~\cite{ErgoregionInstabilityFriedman:1978ygc}. 
The ergotorus is expected to vanish when considering the full field equations with realistic parameters, beyond the thin string approximation, as demonstrated in~\cite{KunzRaduBintoro_PhysRevD.87.104022}. 
Fig.~\ref{fig:ergotoruscross} clearly illustrates that the ergotorus surface diminishes as $G\mu$ decreases. For a GUT-scale string with $G\mu\sim 10^{-6}$, the ergotorus region becomes nearly indistinguishable from the vorton core. At this scale, specifically within a distance less than the core thickness $\delta$ from the vorton core, our weak-field assumption fails. Consequently, the ergotorus can be disregarded, much like how the Schwarzschild radius of the Sun is meaningless.%It can be clearly seen from Fig.~\ref{fig:ergotoruscross} that the ergotorus surface gets smaller as $G\mu$ decreases. For the GUT-scale string of $G\mu\sim 10^{-6}$, the ergotorus region would be virtually indistinguishable from the vorton core itself. At this scale, near the vorton core (or at least less than the core thickness $\delta$), our weak-field assumption breaks down and therefore the ergotorus could be ignored, just like the Schwarzschild radius of the Sun.

\subsection{Angular deficit}

As with any cosmic string loop, the spacetime around a vorton exhibits a conical singularity. The deficit angle around the string core in toroidal coordinates is defined as
\begin{equation}
\label{deficitangletor}
%    \delta\psi=2\pi-\lim_{\sigma_0\rightarrow \infty}\frac{\int_{-\pi}^\pi RN^{-2}e^{\eta-\nu}\big|_{\sigma=\sigma_0}d\psi}{\int_{\sigma_0}^\infty RN^{-2}e^{\eta-\nu}d\sigma},
%\end{equation}
%and can also be written as
%\begin{equation}
    \delta\psi=2\pi-\lim_{\sigma_0\rightarrow \infty}\frac{\int_{-\pi}^\pi N^{-2}e^{C}\big|_{\sigma=\sigma_0}d\psi}{\int_{\sigma_0}^\infty N^{-2}e^{C}d\sigma},
\end{equation}
where $C(\sigma,\psi)\equiv \eta-\nu$. Applying l'Hopital's rule for the limit gives
\begin{equation}
    \delta\psi = 2\pi \lim_{\sigma_0\rightarrow \infty}\frac{\partial C}{\partial \sigma}\bigg|_{\sigma=\sigma_0}.
\end{equation}
In our case, with $\eta=0$, we have $C\to-\nu$. Therefore, using \eqref{nuinf} we obtain
\begin{eqnarray}
%\begin{split}
    \lim_{\sigma_0\rightarrow \infty}\frac{\partial C}{\partial \sigma}\bigg|_{\sigma=\sigma_0}&=&-\frac{\nu(\sigma\rightarrow\infty)}{\partial \sigma}\nonumber\\
       &=&4G\mu,
%\end{split}
\end{eqnarray}
which yields the deficit angle around the circular vorton as
\begin{equation}
\label{deficitanglefinal}
    \delta\psi=8\pi G\mu.
\end{equation}
This result is exactly the same as for the case of a straight string, a straight chiral string \cite{GravityStraightChiralStringSteer:2000jn}, and circular string loops (both static and rotating) \cite{Hughes_PhysRevD.47.468,McManus_PhysRevD.47.1491}.

\section{Gravitational Lensing by Cosmic Vorton}
\label{sec:gravlens}

%The gravitational lensing by string loops in the thin-lens approximation was first studied by de Laix and Vachaspati~\cite{deLaix:1996vc}. They calculated the lensing images produced by a circular loop as well as some non-circular Turok loops with higher frequency modes. With the (weak-field) metric solution in hand, we can do better for the vorton. We calculate the lensing around a circular vorton using the geodesic equations. Assuming that the full Einstein field solutions behave similarly, the solution we obtained should be sufficient to visualize the gravitational lensing around the vorton.
The gravitational lensing by string loops in the thin-lens approximation was first studied by de Laix and Vachaspati~\cite{deLaix:1996vc}, who analyzed the lensing images produced by circular loops as well as non-circular Turok loops with higher-frequency modes. Jusufi investigated the weak deflection of light by a spinning loop within Einstein-Cartan gravity~\cite{Jusufi:2016wiz, Jusufi:2017hed}, suggesting that the deflection angle depends on intrinsic spin and torsion. However, in a subsequent study, Jusufi and Ovg\"un~\cite{Jusufi:2017uhh, Ovgun:2018xys} applied the Gauss-Bonnet theorem and concluded that the deflection angle is independent of spin but is strongly influenced by the cosmological constant $\Lambda$.

With the weak-field metric solution in hand, we can improve upon de Laix-Vachaspati's results for the vorton case. Specifically, we calculate the lensing around a circular vorton using the geodesic equations. Assuming that the full Einstein field solutions behave similarly, the solution we obtained should be sufficient to visualize the gravitational lensing around the vorton. This approach allows for a more accurate and detailed understanding of the lensing effects produced by vortons.

The acceleration of light from the null geodesic equation is given by%}We have obtained the full weak-field metric of the circular chiral vorton. We expect the full Einstein field equation solution to behave the same way, and therefore we assume the solution we obtained is sufficient to make the visualization of the gravitational lensing around the object. The acceleration of light from the geodesic equation is
\begin{equation}
\label{geodeq}
        \frac{d^2x^\mu}{d\lambda^2}=-\Gamma^\mu_{\alpha\beta}\frac{dx^\alpha}{d\lambda}\frac{dx^\beta}{d\lambda}.
\end{equation}
They are:
\begin{eqnarray}
\ddot{t}&=&-2\left(\Gamma^t_{tr}\ \dot{t}\ \dot{r}+\Gamma^t_{\phi r}\ \dot{\phi}\ \dot{r}\right),\nonumber\\
\ddot{\phi}&=&-2\left(\Gamma^{\phi}_{tr}\ \dot{t}\ \dot{r}+\Gamma^{\phi}_{\phi r}\ \dot{\phi}\ \dot{r}\right),\nonumber\\
\ddot{r}&=&-\left(\Gamma^r_{tt}\ \dot{t}^2+2\Gamma^r_{t\phi}\ \dot{t}\ \dot{\phi}+\Gamma^r_{\phi\phi}\ \dot{\phi}^2+\Gamma^r_{rr}\ \dot{r}^2+\Gamma^r_{zz}\ \dot{z}^2\right),\nonumber\\
\ddot{z}&=&-2\Gamma^z_{rz}\ \dot{r}\ \dot{z},
\end{eqnarray}
where $\dot{u}\equiv du/d\lambda$. We also have the constraints
\begin{equation}
g_{tt}\Dot{t}^2+2g_{t\phi}\Dot{t}\Dot{\phi}+\Vec{v}^2=0,
\end{equation}
with
\begin{equation}
\Vec{v}^2\equiv g_{ij}\frac{dx^i}{d\lambda}\frac{dx^j}{d\lambda}.
\end{equation}
We can then express $\Dot{t}$ as
\begin{equation}
\label{tdotsub}
    \Dot{t}=\frac{-g_{t\phi}\Dot{\phi}+\sqrt{g_{t\phi}^2\Dot{\phi}^2-g_{tt}\Vec{v}^2}}{g_{tt}}.
\end{equation}
This expression for $\dot{t}$ can be substituted into Eq.~\eqref{geodeq}, which yields conditions that can be solved using spatial initial conditions alone. Consequently, the required acceleration is reduced to the three-acceleration $\vec{a} = \ddot{x}$ of the null particle.
%This simplification streamlines the analysis, allowing us to focus solely on the spatial dynamics of the particle's motion.}
%This can be inserted into \eqref{geodeq} to form equations solvable by using spatial initial conditions alone. Now, the acceleration needed is only the three-acceleration $\Vec{a}=d^2\Vec{x}/d\lambda^2$ of the null particle. 

The full explicit form of the geodesic equations would be excessively long and unilluminating. However, the key difference from the Schwarzschild configuration lies in the symmetry. Unlike the Schwarzschild case, where we can set $\theta=\pi/2$ without loss of generality, this choice is not valid here, as the metric depends on both $r$ and $z$. The symmetry of the motion is instead more akin to that of the Kerr configuration, with a crucial distinction: in Kerr spacetime, the singularity is enclosed within an event horizon, whereas for the vorton, no event horizon exists.

\begin{figure}[H]
     \centering
     \begin{subfigure}[b]{0.45\textwidth}
         \centering
\includegraphics[width=\textwidth]{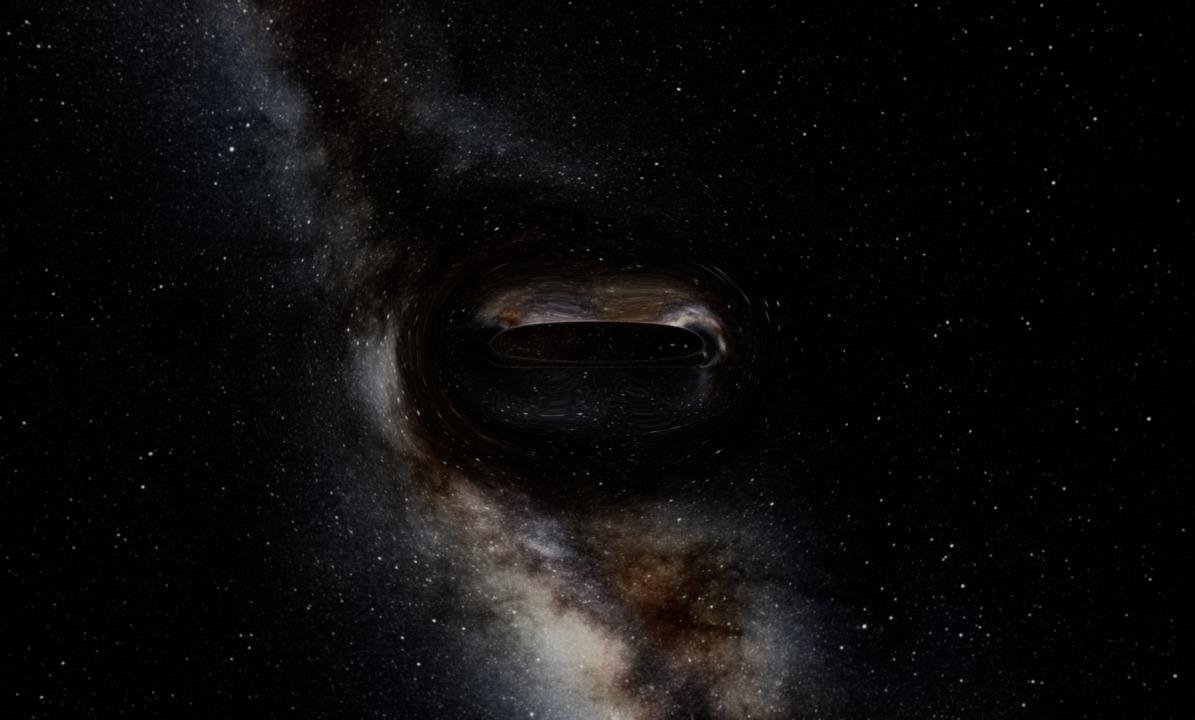}
\caption{$4\pi\mu=0.06$}%$M=10$, $R=50$}
     \end{subfigure}
     \hfill
    \begin{subfigure}[b]{0.45\textwidth}
         \centering
\includegraphics[width=\textwidth]{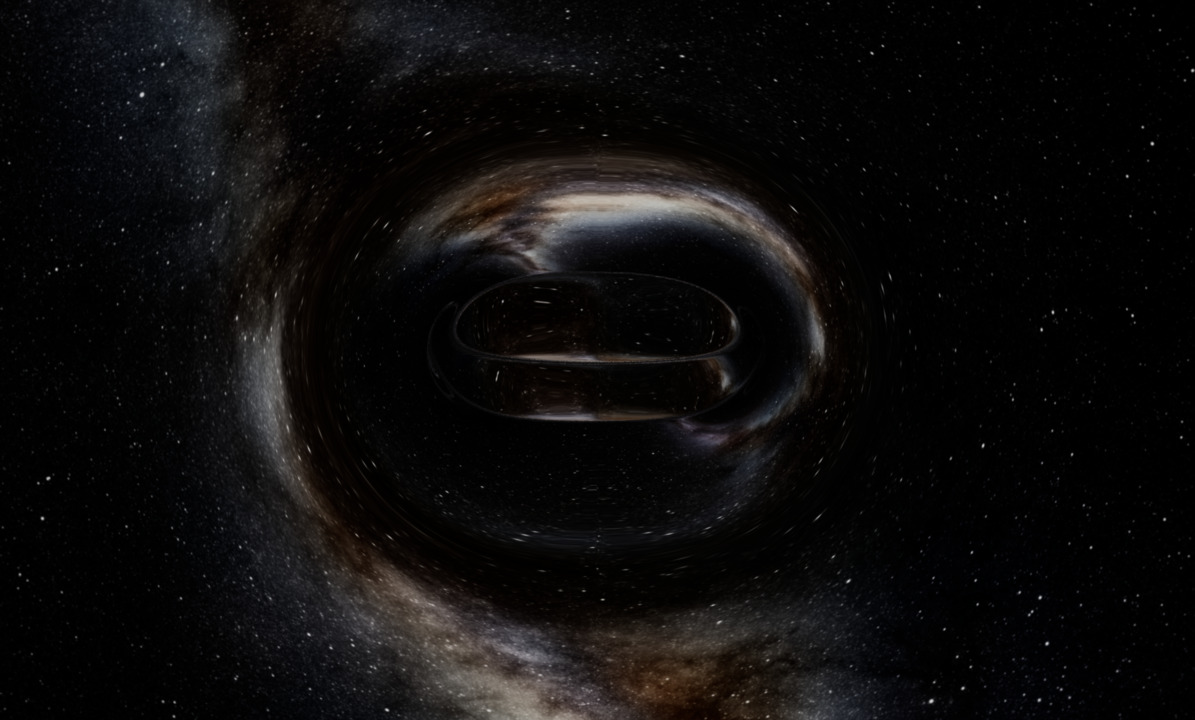}
\caption{$4\pi\mu=0.2$}%$M=10$, $R=100$}
     \end{subfigure}
    \caption{Lensing images by a chiral vorton at $\phi=0.035\pi$.}
    \label{fig:6.3degviz}
\end{figure}

\begin{figure}[H]
     \centering
     \begin{subfigure}[b]{0.4\textwidth}
         \centering
\includegraphics[width=\textwidth]{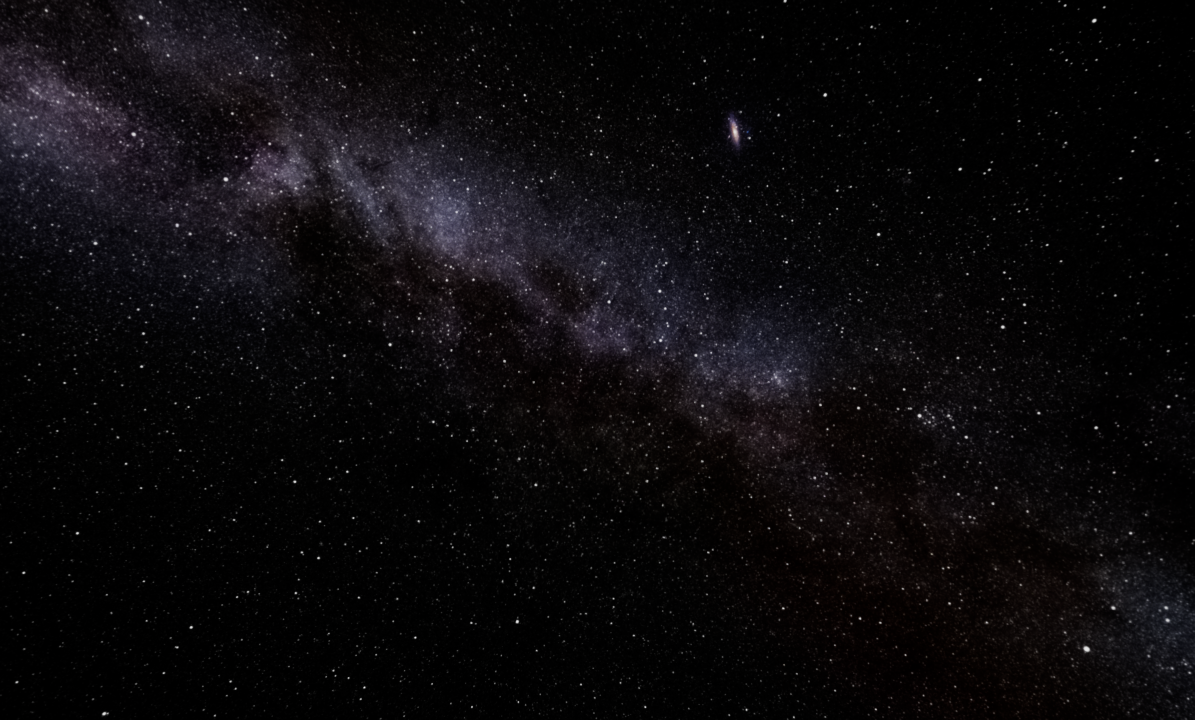}
         \caption{Unlensed}
     \end{subfigure}
     \hfill
     \begin{subfigure}[b]{0.4\textwidth}
         \centering
\includegraphics[width=\textwidth]{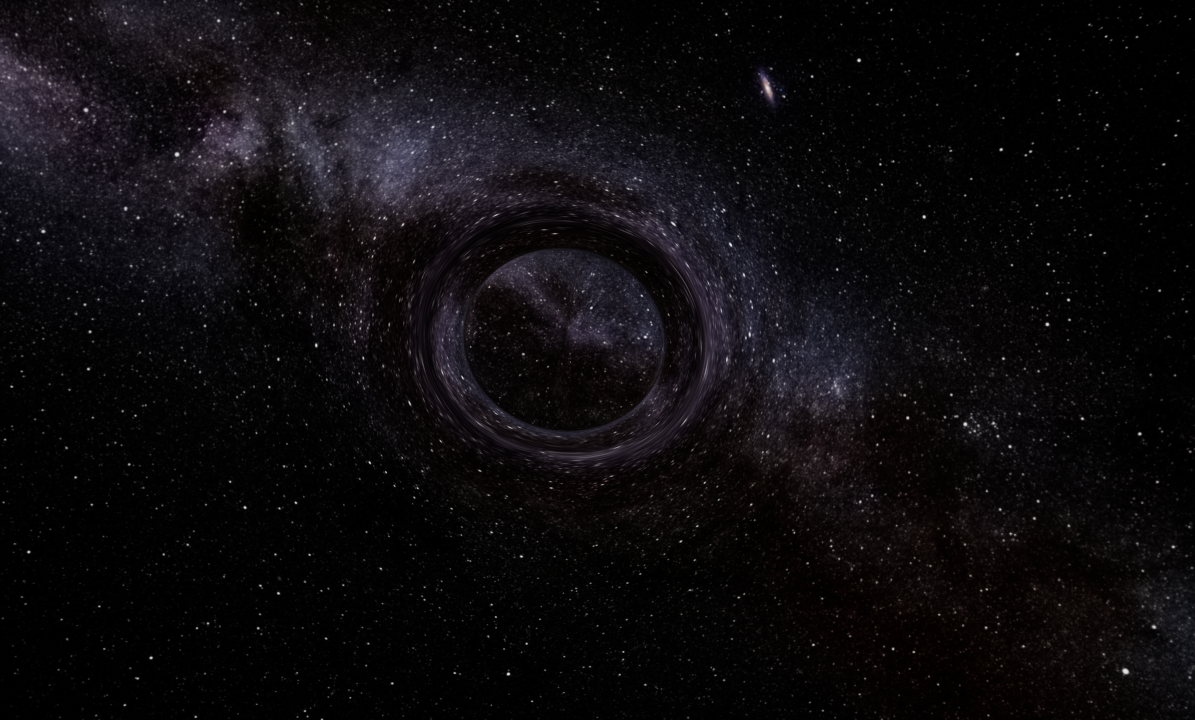}
\caption{$4\pi\mu=0.06$}%$M=3$, $R=50$}
     \end{subfigure}
     \hfill
    \begin{subfigure}[b]{0.4\textwidth}
         \centering
\includegraphics[width=\textwidth]{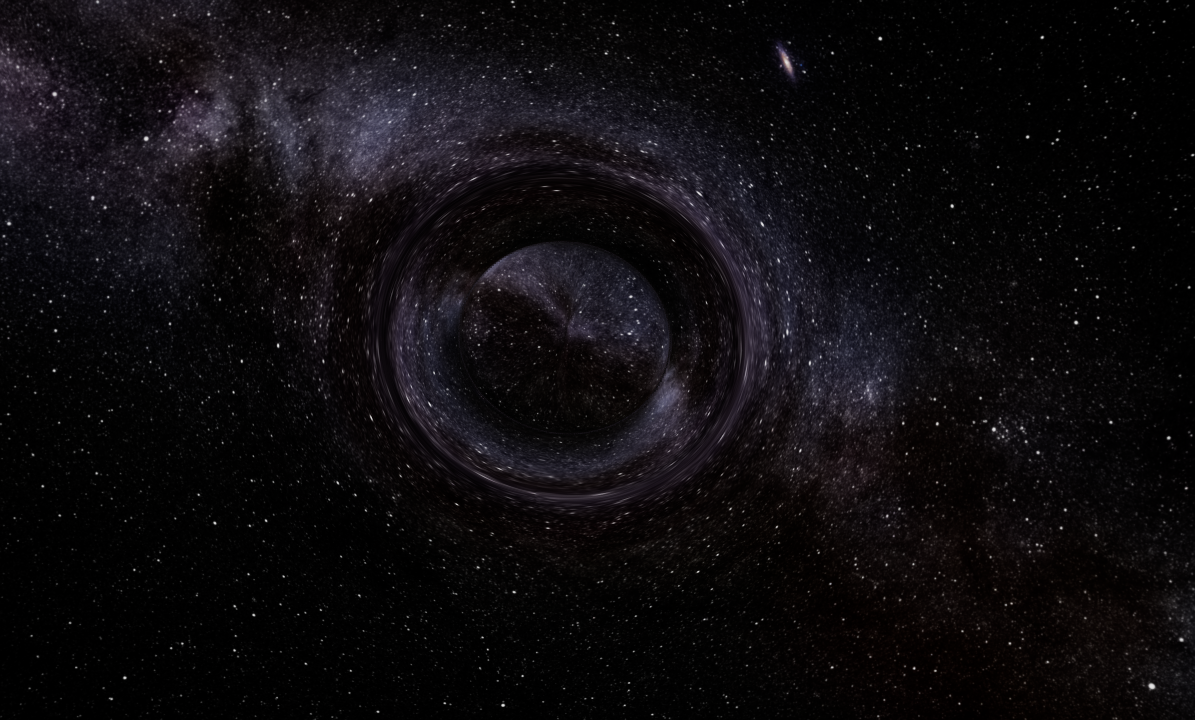}
\caption{$4\pi\mu=0.1$}%$M=10$, $R=100$}
     \end{subfigure}
    \caption{Lensing images by a chiral vorton at $\phi=0.35\pi$.}
        \label{fig:perpviz}
\end{figure}

\begin{figure}[H]
     \centering
     \begin{subfigure}[b]{0.45\textwidth}
         \centering
\includegraphics[width=\textwidth]{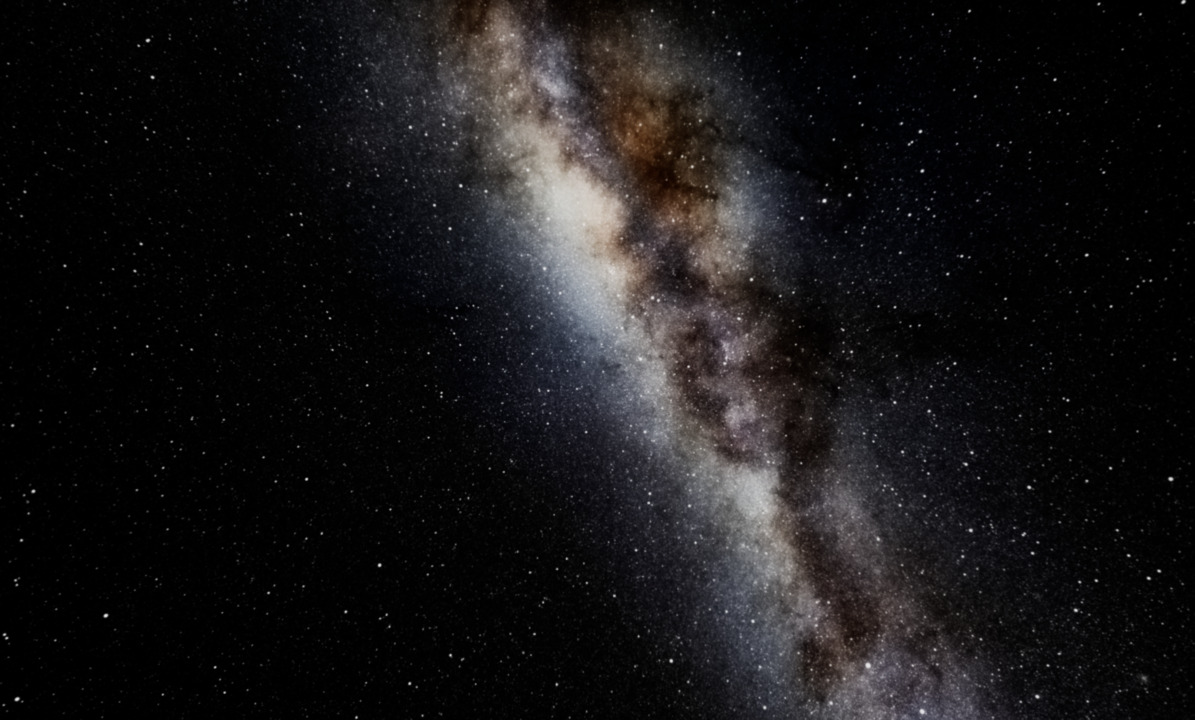}
         \caption{Unlensed}
     \end{subfigure}
     \hfill
     \begin{subfigure}[b]{0.45\textwidth}
         \centering
\includegraphics[width=\textwidth]{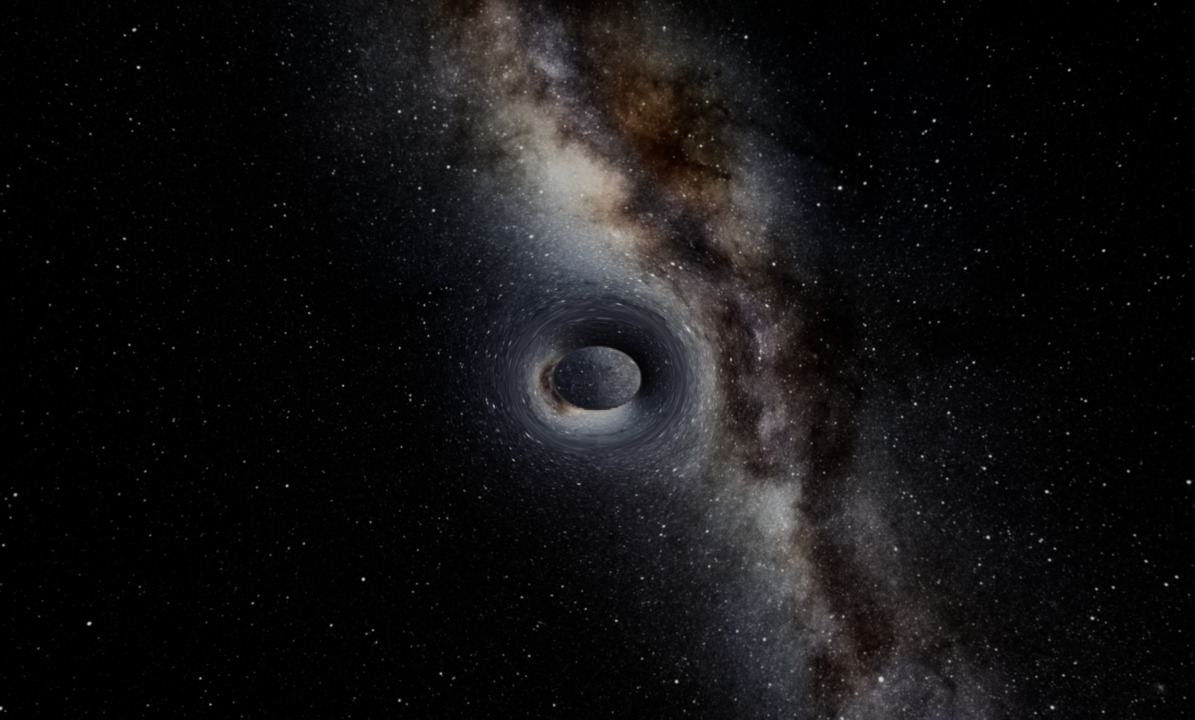}
\caption{$4\pi\mu=0.05$}%$M=1$, $R=20$}
     \end{subfigure}
     \hfill
%     \begin{subfigure}[b]{0.45\textwidth}
%         \centering
%\includegraphics[width=\textwidth]{45DegOffPlanem1r50.jpg}
         %\caption{$M=1$, $R=50$}
%     \end{subfigure}
%     \hfill
    \begin{subfigure}[b]{0.45\textwidth}
         \centering
\includegraphics[width=\textwidth]{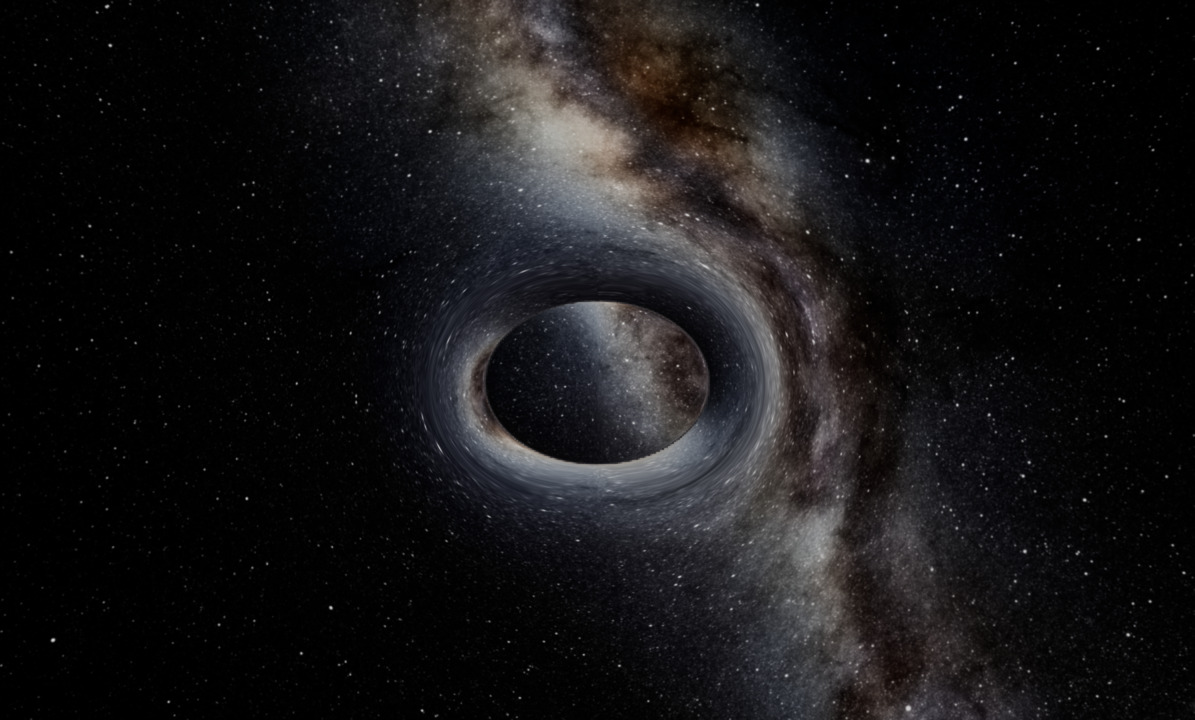}
\caption{$4\pi\mu=0.06$}%$M=3$, $R=50$}
     \end{subfigure}
    \caption{Lensing images by a chiral vorton at $\phi=\pi/4$.}
        \label{fig:45viz}
\end{figure}

\begin{figure}[H]
     \centering
     \begin{subfigure}[b]{0.45\textwidth}
         \centering
\includegraphics[width=\textwidth]{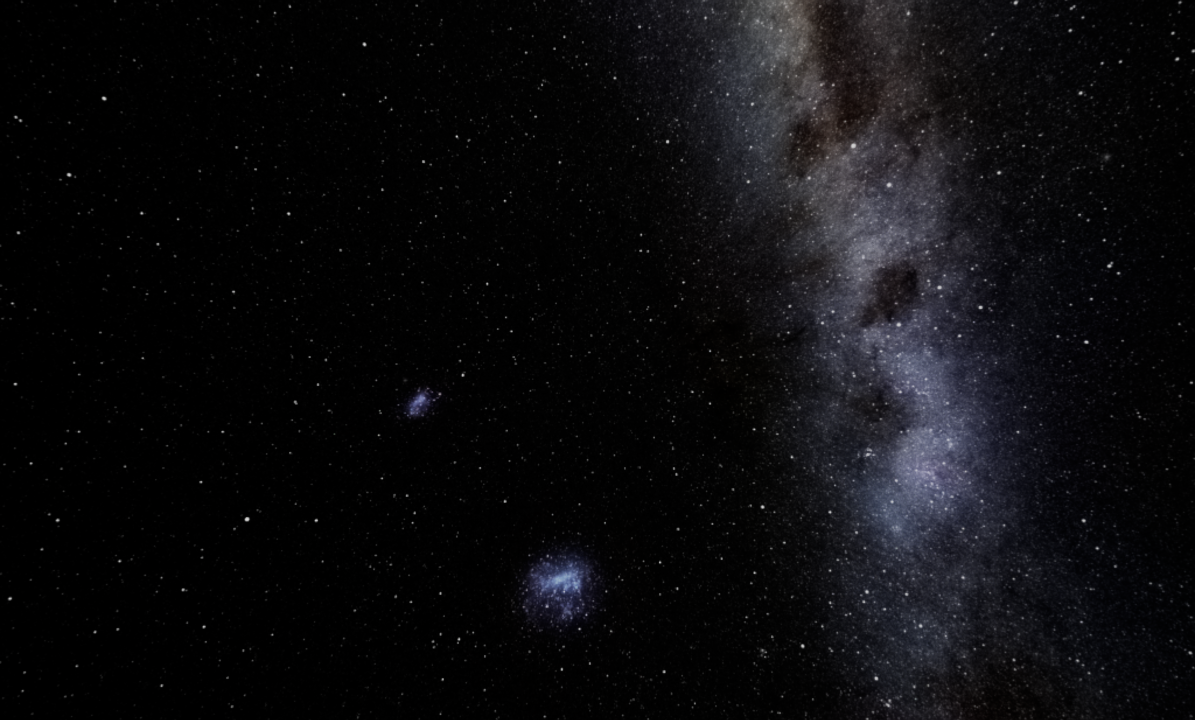}
\caption{$4\pi\mu=0.06$}
     \end{subfigure}
     \hfill
          \begin{subfigure}[b]{0.45\textwidth}
         \centering
\includegraphics[width=\textwidth]{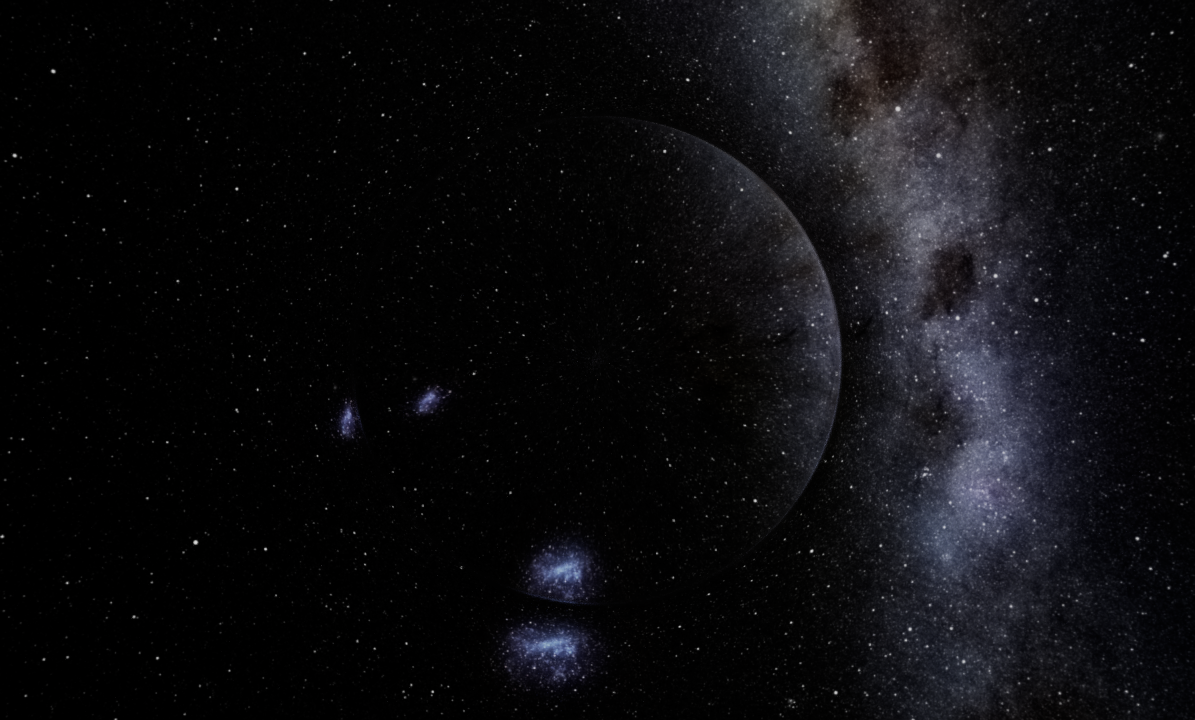}
\caption{$4\pi\mu=0.025$}
     \end{subfigure}
     \hfill
          \begin{subfigure}[b]{0.45\textwidth}
         \centering
\includegraphics[width=\textwidth]{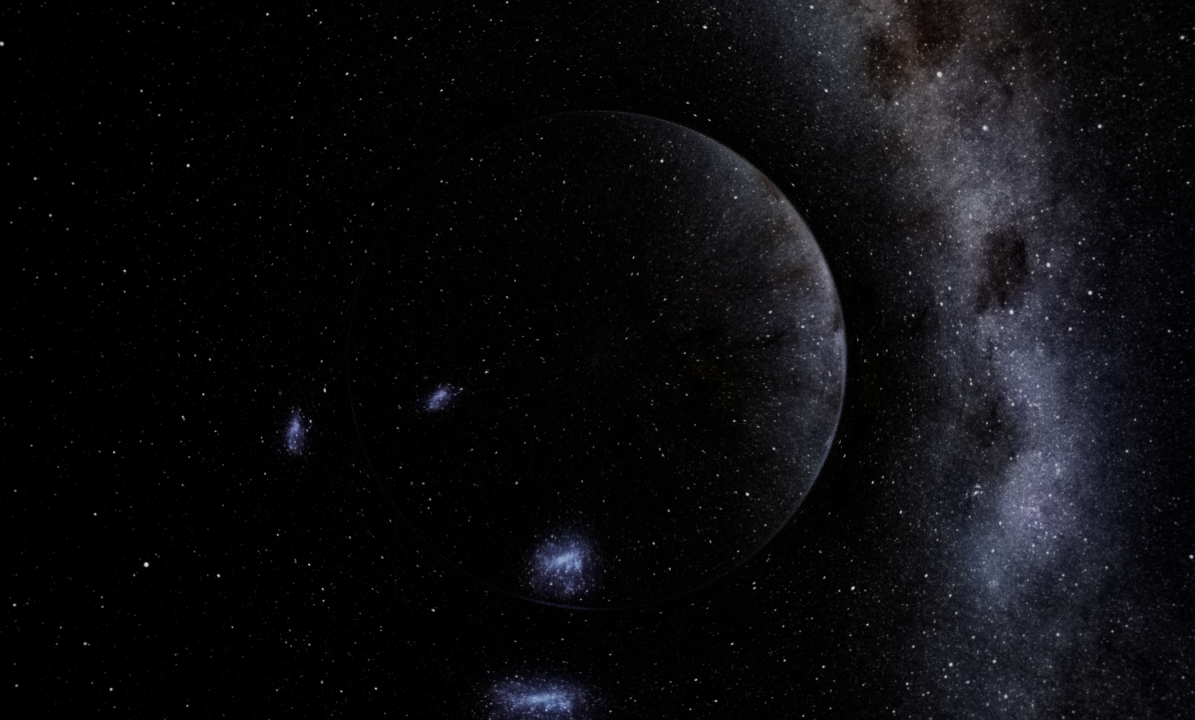}
\caption{$4\pi\mu=0.05$}
     \end{subfigure}
     \hfill
     \begin{subfigure}[b]{0.45\textwidth}
         \centering
\includegraphics[width=\textwidth]{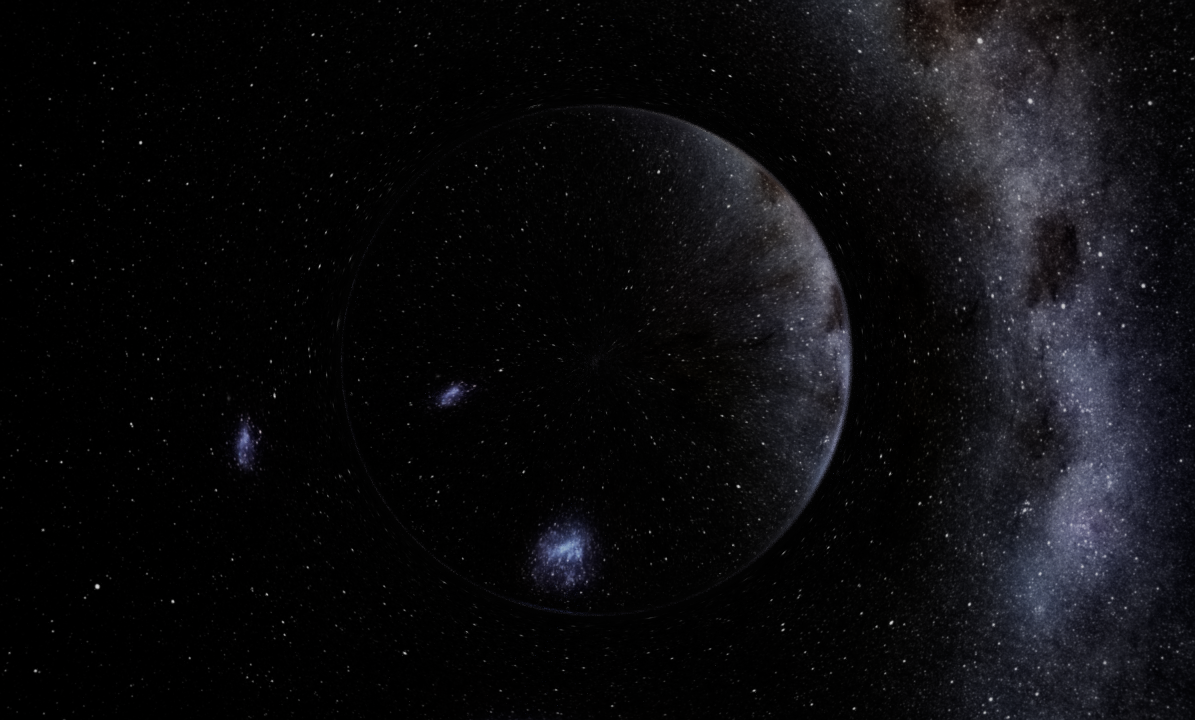}
\caption{$4\pi\mu=0.075$}
     \end{subfigure}
    \caption{Visualization of vorton gravitational lensing from perpendicular point of view ($\phi=\pi/2$).}
\label{fig:90viz}
\end{figure}

\begin{figure}[H]
     \centering
     \begin{subfigure}[b]{0.45\textwidth}
         \centering
\includegraphics[width=\textwidth]{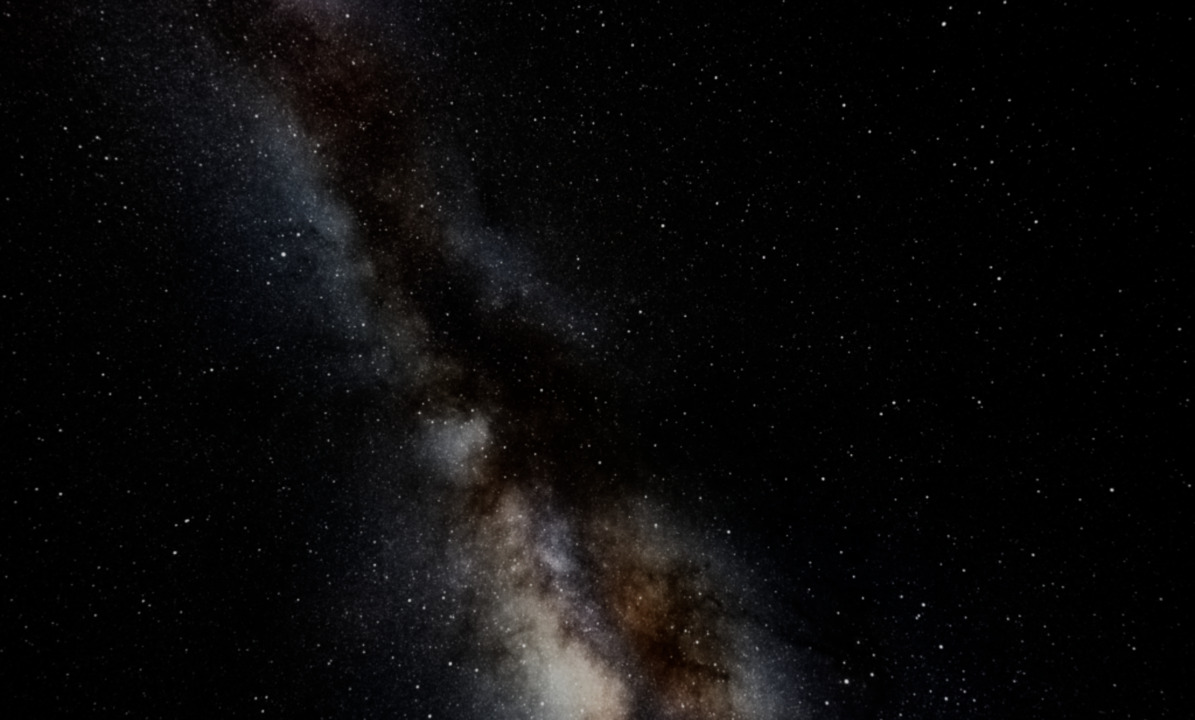}
         \caption{Unlensed}
     \end{subfigure}
     \hfill
    \begin{subfigure}[b]{0.45\textwidth}
         \centering
\includegraphics[width=\textwidth]{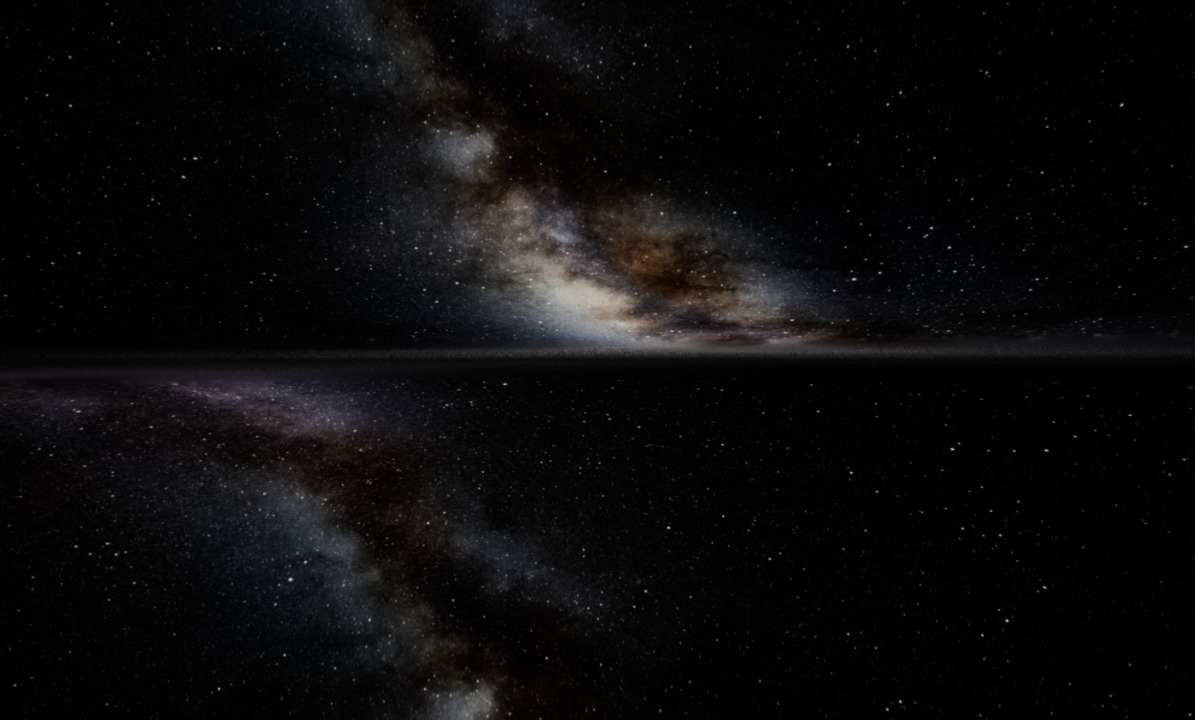}
\caption{$4\pi\mu=0.2$}%$M=10$, $R=50$}
     \end{subfigure}
     \hfill
    \begin{subfigure}[b]{0.45\textwidth}
         \centering
\includegraphics[width=\textwidth]{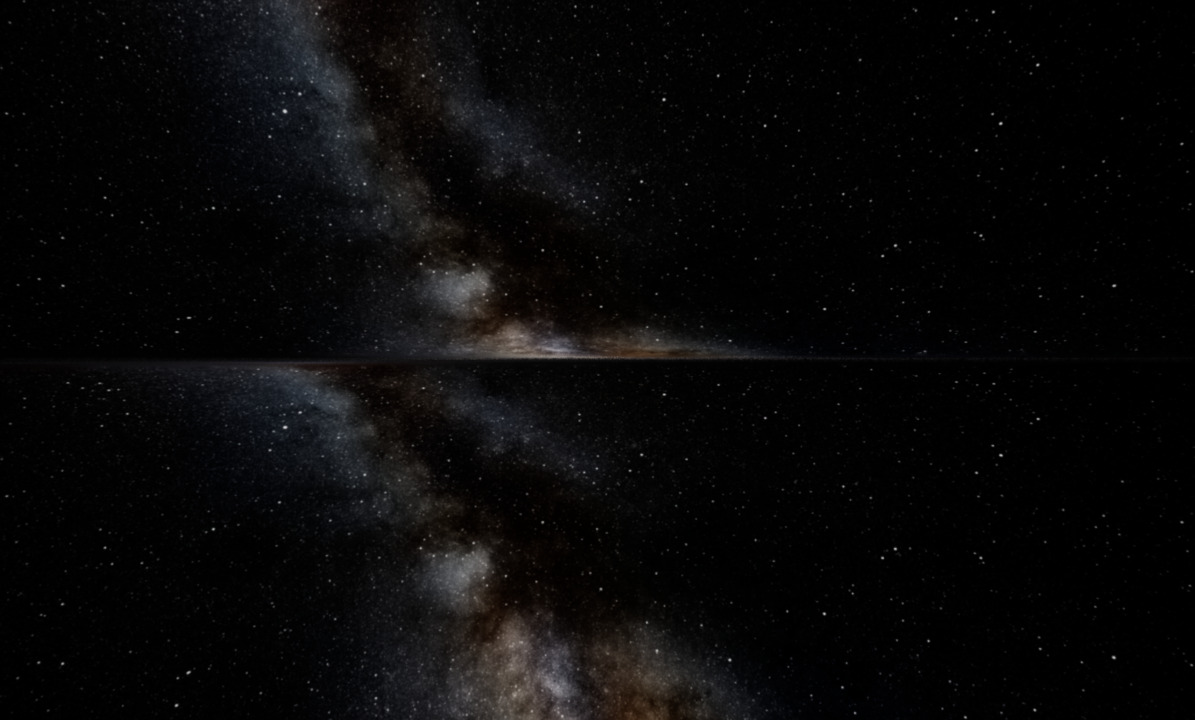}
\caption{$4\pi\mu=0.1$}%$M=10$, $R=100$}
     \end{subfigure}
\caption{Lensing images by a chiral vorton from the center of the loop.}
\label{fig:centralviz}
\end{figure}

\begin{figure}[H]
     \centering
     \begin{subfigure}[b]{0.45\textwidth}
         \centering
\includegraphics[width=\textwidth]{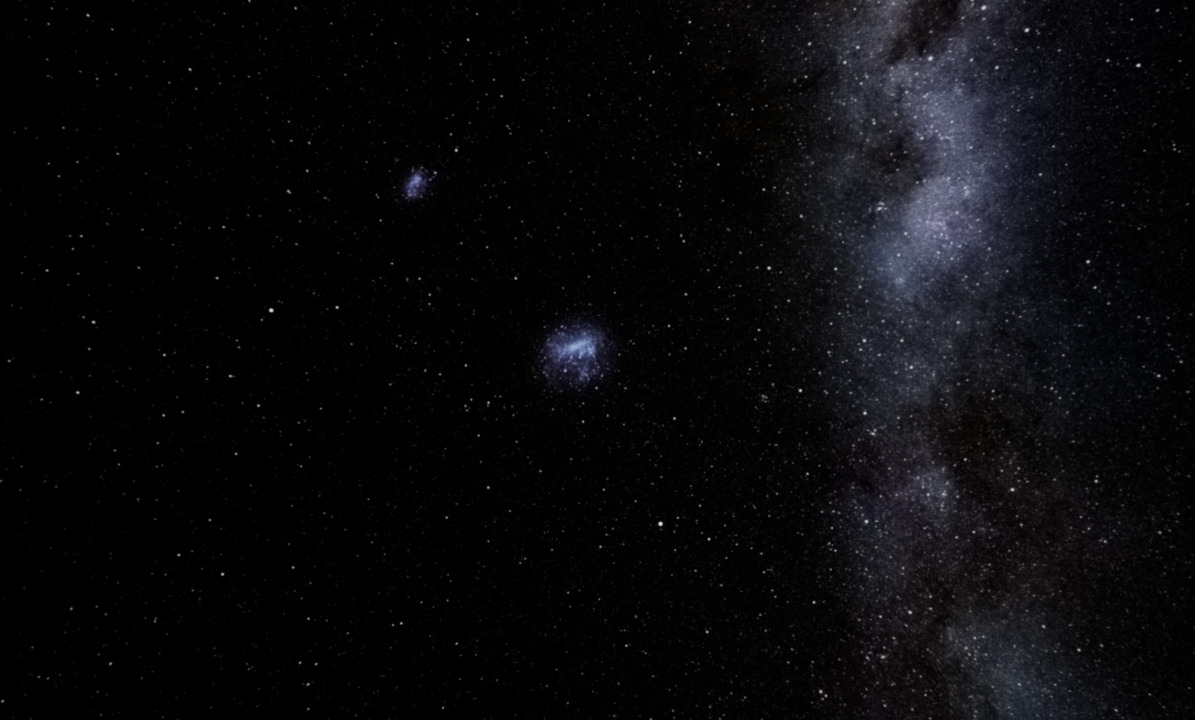}
         \caption{Unlensed}
     \end{subfigure}
     \hfill
     \begin{subfigure}[b]{0.45\textwidth}
         \centering
\includegraphics[width=\textwidth]{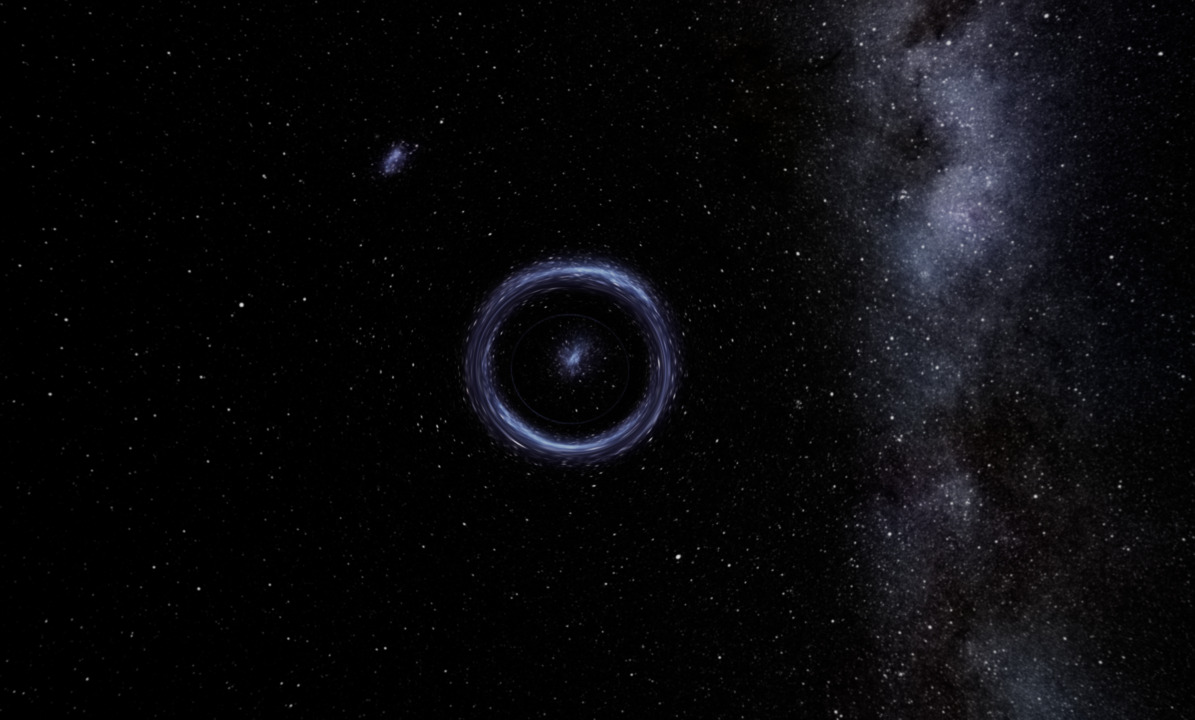}
\caption{$4\pi\mu=0.02$}%$M=1$, $R=20$}
     \end{subfigure}
     \hfill
     \begin{subfigure}[b]{0.45\textwidth}
         \centering
\includegraphics[width=\textwidth]{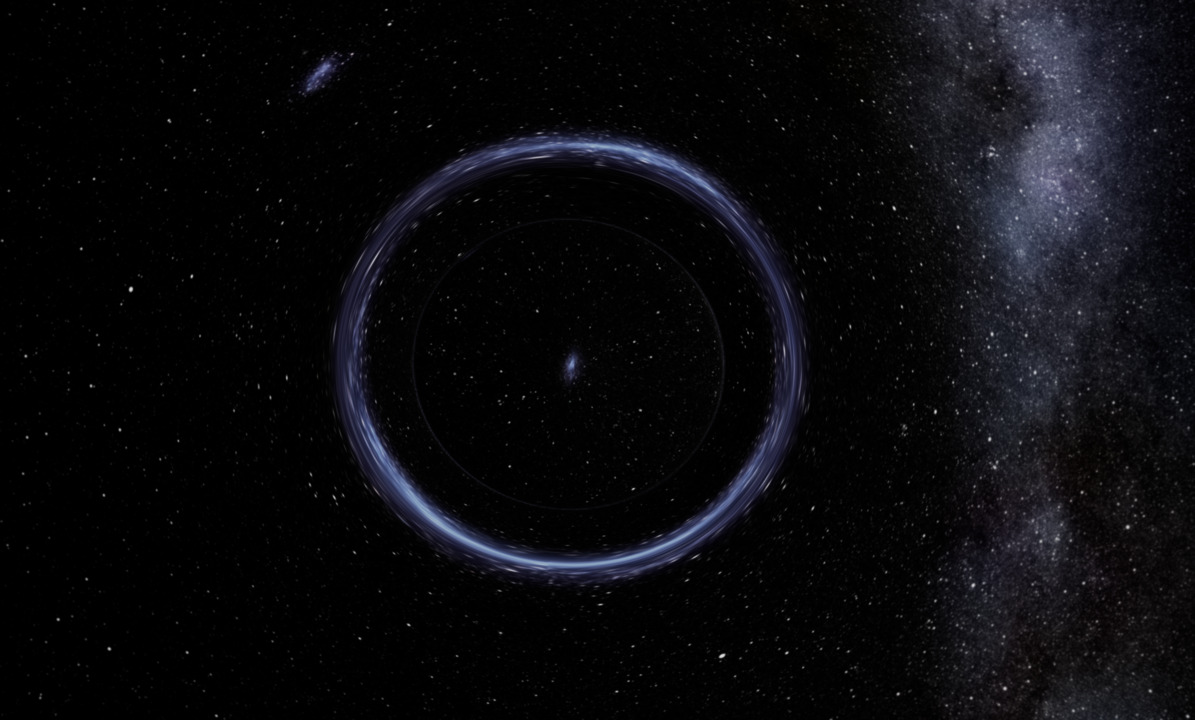}
\caption{$M=3$, $R=50$}
     \end{subfigure}
    \caption{Visualization of Einstein ring together with the weakly distorted source inside the loop generated by vorton.}
\label{fig:eringviz}
\end{figure}

The lensing calculation was performed numerically. To do so, we need an analytic approximation of the functions to simplify the calculations. %To facilitate this, a celestial sphere was employed in the scene setup. This celestial sphere, an inverted sphere with a celestial texture cubemap assigned to it, utilized the deep starmap \cite{NASAStarmap} as its cubemap texture. Positioned at the center of the sphere, the vorton was represented by an image derived from the circular (chiral) vorton metric obtained previously. In this setup, the celestial sphere boasted a radius of $5000$ units. 
Our results are shown in Figs.~\ref{fig:6.3degviz}-\ref{fig:eringviz}. We show lensing images for various tension and angle $\phi$ (the angle between the vorton's plane and the detector's axis). We also provide the unlensed image for comparison.%The visualization is done using Unreal Engine 5, which requires an analytic approximation of the functions. The scene setup uses a celestial sphere, which is an inverted sphere with celestial texture cubemap assigned on it. Here we use the deep starmap \cite{NASAStarmap} as the cubemap texture. The celestial sphere has the radius of $5000$ units with the vorton in the center of the sphere. The image of circular (chiral) vorton from the metric obtained before is rendered in Unreal Engine 5 with the following results.

The lensing images produced by a circular vorton vary significantly with different angles $\phi$, representing potential detections from various observational viewpoints.  Figs.~\ref{fig:6.3degviz}, \ref{fig:perpviz}, \ref{fig:45viz}, and \ref{fig:90viz} illustrate typical image distortions as $\phi$ increases up to $\pi/2$. Fig.~\ref{fig:centralviz} illustrates the image detected when the detector is positioned at the center of the vorton loop. Fig.~\ref{fig:eringviz} depicts an Einstein ring, occurring when the source, vorton loop, and detector are aligned. As $\mu$ increases, image distortion becomes more pronounced.

For an observer viewing a perpendicular circular vorton (Fig.~\ref{fig:90viz}), there would be a weakly distorted image inside the string, and a highly distorted image, or even an Einstein ring (Fig.~\ref{fig:eringviz}), outside the string, similar to that in \cite{deLaix:1996vc}, though it was obtained using different approach and different scale. At every angle, there is an apparent ring-shaped discontinuity in the image, a generic feature of cosmic strings due to their conical geometry. The discontinuity of image in this case is the manifestation of the deficit angle of the string. Furthermore, from the observer perpendicular to the vorton plane, for example. in Figs.~\ref{fig:perpviz}, we see no apparent difference between the lensing pattern to that of Kerr black hole. However, an observer parallel (or at least nearly parallel) to the vorton plane will notice the difference between the vorton lensing pattern and that of Kerr black hole, for example in \cite{KerrLensingPhysRevD.101.044031,James_2015}. The vorton lensing on the vorton plane, for example in Fig.~\ref{fig:6.3degviz}, still has a symmetry and an oval shape around the discontinuity, while the Kerr black hole has an asymmetry where the pattern protrudes to one side. %We could also see that in every angle there would be an apparent ring-shaped discontinuity of the image, which is a generic feature of cosmic string that stem from its conical nature of spacetime. The image of a weakly distorted source at the center of an Einstein ring of that same source together with an apparent discontinuity of the image can be the signature of cosmic string loop lensing, which beg for the possibility of observation.%, as it is the case in \cite{arxivsafonova2023deep}.
 The combination of a weakly distorted source at the center of an Einstein ring and the image discontinuity can be a signature of cosmic string loop lensing, suggesting the possibility of observation.
This is true since typically the straight path to the source of an Einstein ring would be blocked by the lens (usually a galaxy) if not distorted to form the ring. Additionally, if the observer views the string from near the loop plane, a double image of the loop's back appears, one on top and one on the bottom, similar to the double images of an accretion disk around a black hole. %normally the straight path to the source of an Einstein ring would be blocked by the lens (usually a galaxy) had it not been distorted to form the ring. We also note that if the observer is seeing the string from near the loop plane, a double image of the back of the loop would appear, one on top and one on the bottom. This interesting effect is akin to that of double images of accretion disc around a black hole. We could also see that from the loop's center (Fig.~\ref{fig:centralviz}), the outside image looks like a double image of weakly lensed image source, with a discontinuity in the middle. This is actually the characteristics of the straight string image, which means that a circular vorton seen from the center of the loop looks like a straight string.
From the loop's center (Fig.~\ref{fig:centralviz}), the outside image appears as a double image of a weakly-lensed source with a central discontinuity, resembling the image of a straight string. This indicates that a circular vorton seen from the center of the loop looks like a straight string. 

\section{Conclusions}
\label{sec:conc}

Cosmic strings are among the most extensively studied topological defects in cosmology, as their dynamics can evade the overclosure of the universe. Among various string loop configurations, vortons are theoretically considered the most stable, with their oscillations balanced by angular momentum. This remarkable stability raises concerns that vortons may not be diluted, potentially leading to an abundance problem. However, some proposals suggest that stable vortons could instead serve as viable dark matter candidates~\cite{Auclair:2020wse}.  

In this paper, we  study the gravitational field and lensing of a circular chiral vorton. Using a weak-field approximation, we solve the corresponding Einstein field equations. The metric solutions $\nu(r,z)$ and $A(r,z)$ are analogous to the $A_t$ and $A_{\varphi}$ components, respectively, of the vector potential of a circular current wire loop. Around the string core, the spacetime exhibits conical singularity with deficit angle $\delta\psi=8\pi G\mu$, as in the straight string. There exists ergosurface with $S^1\times S^1$ topology, dubbed the {\it ergotorus}, between the ergoregion and the external space. The appearance of the ergotorus 
 indicates instability, which we expect to be smoothed out in the thick string approach.

As expected, the field diverges at the vorton's core. To understand the origin of the divergence, note that we employ two approximations: the weak-field limit of gravity and the thin-string approximation. The former means that Einstein’s equations are truncated at first order, while the latter treats the string core as a delta-function source, as in Eqs.~\eqref{16}-\eqref{17}. It is the property of the delta function that gives rise to the singularity at the core. One might ask whether relaxing the weak-field approximation could smooth out the singularity. The answer is no. As long as the thin-string approximation remains in effect, the singularity persists. It only disappears when the core is resolved as a thick string, as demonstrated in~\cite{KunzRaduBintoro_PhysRevD.87.104022}.
 
Asymptotically, the spacetime around a vorton loop of radius 
$R$ is identical to a Kerr black hole with mass $M_v=4\pi R\mu$, spin parameter $a=R/2$, and angular momentum $J_v=2\pi R^2\mu$. This is intriguing because what we previously detected as a distant rotating black hole might actually be a vorton. More surprisingly, for a typical GUT-scale string ($R_v\sim10^5-10^6R_S$), Eq.~\eqref{nakedsingularitycondition} is always satisfied, saturating the extremal bound for Kerr. Thus, a GUT-scale vorton is indistinguishable from a Kerr naked singularity to a distant observer. It is worth noting that recently there has been a growing interest in the study of Kerr naked singularities, spurred by the possible existence of Kerr superspinars within the string theory framework~\cite{Gimon:2007ur}. Researchers are particularly focused on photon orbits~\cite{Charbulak:2018wzb, NooriGashti:2024gnc} and the shadows~\cite{Nguyen:2023clb,Mangut:2023oxa} of Kerr naked singularities.  Studying the null and timelike geodesics around a vorton could provide insights into distinguishing it from a Kerr naked singularity.

Recently, a suspected gravitational lensing detection of a cosmic string has been reported~\cite{arxivsafonova2023deep}. Using photometric and spectroscopic observations, researchers propose that the galaxy pair SDSSJ110429 are actually duplicate images of a single galaxy lensed by a cosmic string. Given that a vorton is postulated to be a stable string loop configuration, our study of lensing images by a vorton becomes particularly relevant in this context. Our study reveals unique properties of vorton-generated lensing images. They have an apparent discontinuity that separates minimally distorted regions from highly distorted ones from the same source, as can be seen, for example, in Figs.~\ref{fig:perpviz}-\ref{fig:45viz}. Moreover, there are similarities between the lensing images due to vorton and a black hole with an accretion disk. This can be observed, for example in Fig.~\ref{fig:6.3degviz}, where the ring image could form twice simultaneously, with light from both the upper and lower sides of the ring being deflected back towards a camera positioned near the plane of the vorton. An Einstein ring could form around a vorton, while the original image of the source remains visible with minimal distortion at the center of the ring, as shown in Fig.~\ref{fig:eringviz}. Typically, an Einstein ring forms when another object, which creates the ring, blocks the direct view of the source image, distorting it into the ring shape. The simultaneous appearance of both the ring and the almost undistorted source can be a distinctive signature of a string loop in the sky.

This preliminary work opens up possibilities for studying the general gravitational lensing and dynamics of vortons. Here, we focus on the geometry and geodesics of a circular chiral vorton, as this loop shape possesses the highest symmetry, making the metric solution analytically tractable. Other loop shape configurations, reminiscent of the Kibble-Turok two-parameter family of solutions~\cite{Kibble:1982cb, Turok:1984cn} in the case of Nambu-Goto cosmic string loops, certainly exist. We will discuss the gravitational lensing in the thin-lens approximation by arbitrary vorton loop shapes in our next publication. Another issue that can be addressed is the dynamics of vorton loop around a massive object, for example a supermassive black hole. The dynamics of a charged string loop around Reinssner-Nordstrom black hole was discussed in~\cite{CCSLRNOteev:2018gth}. In~\cite{Jacobson:2008cw}, the authors study the dynamics of a current-carrying string loop around a Kerr black hole, suggesting that such a system might help solve the problem of collimated jets in astrophysics. It is interesting to consider the effects of using a vorton instead. Further research is also needed on the dynamics of light and massive objects around a vorton loop. We shall explore these possibilities in subsequent publications.
%\textcolor{blue}{Since vorton is stabilized by its angular momentum, there is frame-dragging effect not present in other string loop configuration. This effect breaks the symmetry of the image}

\acknowledgments

We thank Jose Blanco-Pillado, Steven Holme, and Akif Pervez for the enlightening discussions and constructive suggestions. This work is funded by the Hibah Riset FMIPA UI No.~PKS-042/UN2.F3.D/PPM.00.02/2024. %and Hibah Matching Fund UI-UNTAN No.~PKS-002/UN2.F3.D/PPM.00.02/2024.

\begin{appendix}

\section{Toroidal Coordinate System}
\label{Tor}
\begin{figure}[h]
\centering
\includegraphics[width=0.70\textwidth]
{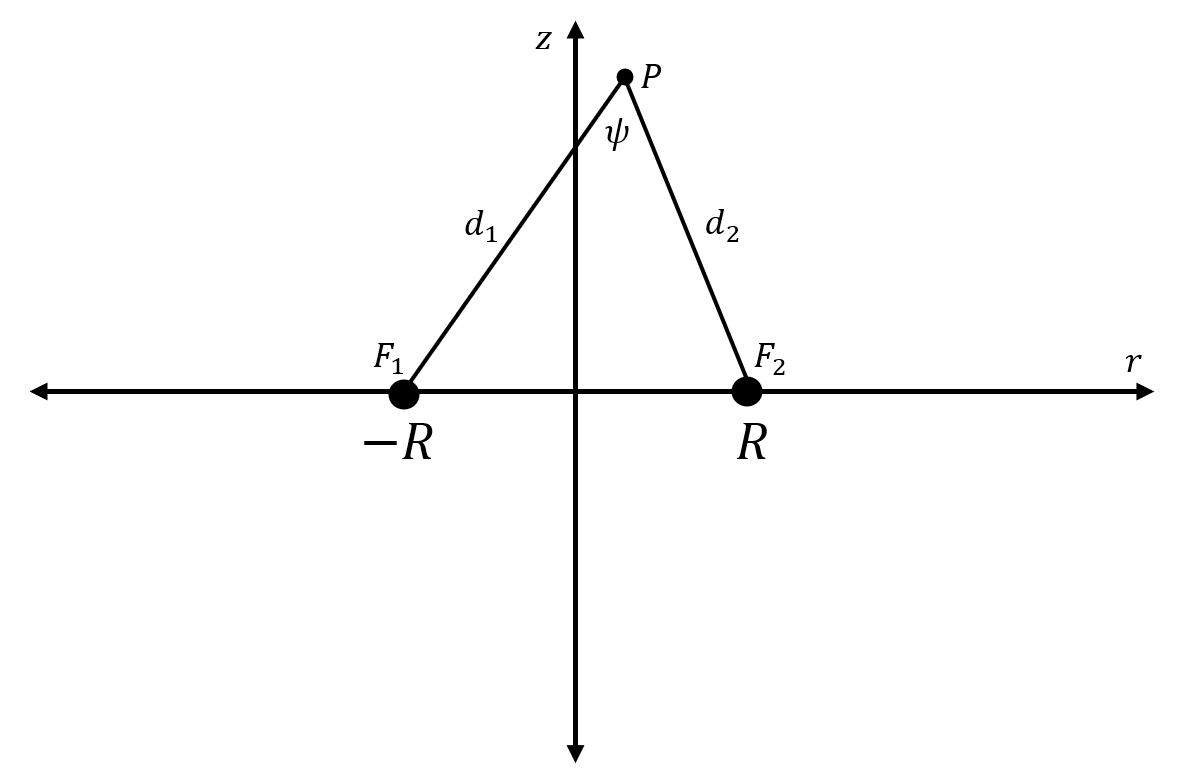}
\caption{Point $P$ in toroidal coordinate.}
\label{fig:ToroidalCoordDef}
\end{figure}

\begin{figure}[h]
\centering
\includegraphics[width=0.70\textwidth]
{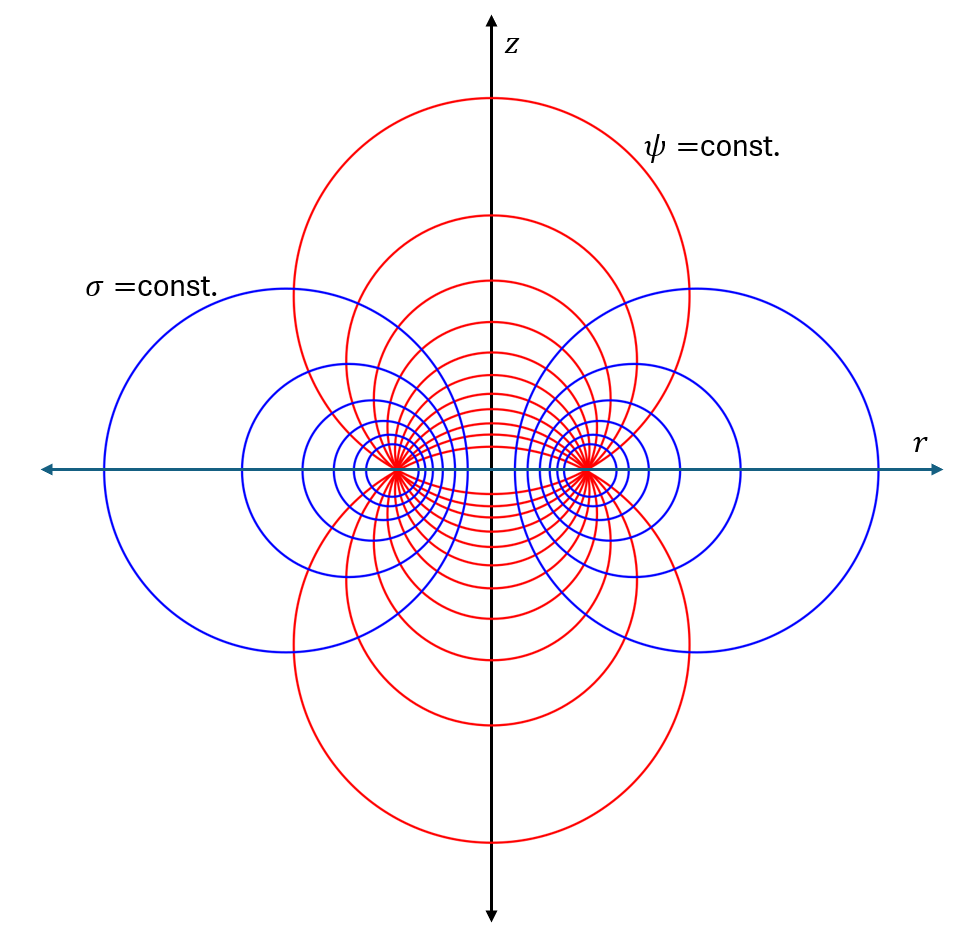}
\caption{Point $P$ in toroidal coordinate.}
\label{fig:ToroidalCoordConstCurve}
\end{figure}

The toroidal coordinate is the coordinate $(\sigma,\psi,\phi)$ defined as follows: Suppose we have a point $P$ cylindrical coordinate $(r,z,\phi)$. We can define two foci $F_1$ and $F_2$ at $(r,z)=(-R,0)$ and $(r,z)=(R,0)$, respectively. If we draw a line from $F_1$ to $P$ and from $F_2$ to $P$, we will get the diagram in Fig. \ref{fig:ToroidalCoordDef}.
\\
The coordinate $\psi$ is the (inner) angle $\angle F_1PF_2$, and if we define $d_1\equiv F_1P$ and $d_2=F_2P$ then the coordinate $\sigma$ is defined as
\begin{eqnarray}
    \sigma\equiv\ln{\frac{d_1}{d_2}}.
\end{eqnarray}
The coordinate transformation in Cartesian coordinate is then
\begin{eqnarray}
    x&=&R\frac{\sinh{\sigma}}{N(\sigma,\psi)^2}\cos{\phi},\\
    y&=&R\frac{\sinh{\sigma}}{N(\sigma,\psi)^2}\sin{\phi},\\
    z&=&R\frac{\sin{\psi}}{N(\sigma,\psi)^2},
\end{eqnarray}
where
\begin{eqnarray} N(\sigma,\psi)^2=\cosh{\sigma}-\cos{\psi}.
\end{eqnarray}
We can see that the curve of constant $\psi$ is in the form
\begin{eqnarray}
    r^2+\left(z-R\cot{\psi}\right)^2=\left(\frac{R}{\sin{\psi}}\right)^2,
\end{eqnarray}
and the curve of constant $\sigma$ is as follows
\begin{eqnarray}
    \left(r-R\coth{\sigma}\right)^2+z^2=\left(\frac{R}{\sinh{\sigma}}\right)^2.
\end{eqnarray}
Clearly, the curves of constant $\psi$ are circles of radius $R/\sin{\psi}$ with center at $(r,z)=(0,R\cot{\psi})$. Meanwhile, the curves of constant $\sigma$ are circles of radius $R/\sinh{\sigma}$ with center at $(r,z)=(R\coth{\sigma},0$. Together, they form what is called the Apollonian circles, that is the two families of circles intersect each other orthogonally, as can be seen in Fig. \ref{fig:ToroidalCoordConstCurve}.

\section{Derivation of $A(r,z)$}
\label{DerA}

The solution of $W(r,z)$ is just the Green's function of operator
\begin{equation}
    \hat{L}\equiv\nabla^2-\frac{1}{r^2},
\end{equation}
in cylindrical coordinate, which can be solved by the elementary method using Fourier transform.  Multiplying \eqref{weq} with an arbitrary  function $f(\theta)$ gives
\begin{equation}
    f(\phi)\nabla^2W(r,z)-\frac{1}{r^2}W(r,z)f(\phi)=-16\pi G\mu\delta(r-R)\delta(z)f(\phi).
\end{equation}
We would like to find a function $f(\phi)$, such that 
\begin{equation}
    f(\phi)\nabla^2W(r,z)-\frac{1}{r^2}W(r,z)f(\phi)=\nabla^2\left(W(r,z)f(\phi)\right).
\end{equation}
Therefore,
\begin{equation}
    \frac{d^2f(\phi)}{d\phi^2}+f(\phi)=0,\ \ \rightarrow\ \ \ f(\phi)=C_1 \cos{\phi}+C_2 \sin{\phi},
\end{equation}
%which have solution in the form
%\begin{equation}
%    f(\phi)=C_1 \cos{\phi}+C_2 \sin{\phi},
%\end{equation}
$C_1$ and $C_2$ being two integration constants which can be set %where we have the freedom to choose $C_1$ and $C_2$, as they would be canceled in the equation anyway. 
%For our purpose, we choose 
 $C_1=1$ and $C_2=0$.
Thus,% we know that
\begin{equation}
\nabla^2\left(W(r,z)\cos{\phi}\right)=\left(\nabla^2W-\frac{W}{r^2}\right)\cos{\phi}.
\end{equation}
Substituting \ref{weq}, and redefining
\begin{equation}
\label{defv}V(r,z,\phi)=W(r,z)\cos{\phi},
\end{equation}
yields
\begin{equation}
    \nabla^2V(r,z,\phi)=-16\pi G\mu\delta(r-R)\delta(z)\cos{\phi}.
\end{equation}
This is the Poisson's equation for the electrostatic potential of a ring charge distribution with radius $R$ centered at $z=0$, with azimuthal charge distribution of $\cos{\phi}$. Using the Green's function for the Laplacian $G(\Vec{r},\Vec{r}')=1/4\pi|\Vec{r}-\Vec{r}'|$, the solution is
\begin{equation}
    V(r,z,\phi)=-\int_R d^3\Vec{r}' \frac{\rho(\Vec{r}')}{4\pi |\Vec{r}-\Vec{r}'|},
\end{equation}
where
\begin{equation}
    \rho(\Vec{r}')=-16\pi G\mu\delta(r'-R)\delta(z')\cos{\phi'}.
\end{equation}
The point $P$ at which the potential is measured has the position vector
\begin{equation}
    \Vec{r}\equiv\Vec{z}+\Vec{\rho},
\end{equation}
where $\Vec{\rho}=r\hat{\rho}$. Suppose $\hat{\rho}$ make an angle $\phi$ with the $x$ axis, and $\hat{r}'$ make an angle $\phi'$ with the $x$ axis. Then,
\begin{eqnarray}
%\begin{split}
    |\Vec{r}-\Vec{r}'|^2&=&|\Vec{z}+\Vec{\rho}-\Vec{r}'|^2\nonumber\\
    &=&z^2+\left(\Vec{\rho}-\Vec{r}'\right)^2\nonumber\\
    &=&z^2+r^2+r'^2-2rr'\cos{(\phi'-\phi)},
%\end{split}
\end{eqnarray}
and the potential becomes
\begin{eqnarray}
%\begin{split}
    V(r,z,\phi)&=&16\pi G\mu \int\int\int r'dr'dz'd\phi' \frac{\delta(r'-R)\delta(z')\cos{\phi'}}{4\pi \sqrt{z^2+r^2+r'^2-2rr'\cos{(\phi'-\phi)}}}\nonumber\\
    &=&4G\mu R\int_0^{2\pi}d\phi' \frac{\cos{\phi'}}{\sqrt{z^2+r^2+R^2-2rR\cos{(\phi'-\phi)}}}.
%\end{split}
\end{eqnarray}
Defining
\begin{equation}
    \phi'=\pi+\phi+2\theta\ \ \ \rightarrow\ \ \  d\phi'=2d\theta,
\end{equation}
%and using $\cos{2\theta}=1-2\sin^2{\theta}$, 
we have
\begin{eqnarray}
%\begin{split}
    V(r,z,\phi)&=&8G\mu R\int_0^{\pi}d\theta \frac{\cos{(\pi+\phi+2\theta)}}{\sqrt{z^2+(r+R)^2-4rR\sin^2{\theta}}}\nonumber\\
    &=&-8G\mu R\bigg(\cos{\phi}\int_0^{\pi}d\theta \frac{\cos{(2\theta)}}{\sqrt{z^2+(r+R)^2-4rR\sin^2{\theta}}}\nonumber\\%\right.\\
    %&\left.
    &&-\sin{\phi}\int_0^{\pi}d\theta \frac{\sin{(2\theta)}}{\sqrt{z^2+(r+R)^2-4rR\sin^2{\theta}}}\bigg).
%\end{split}
\end{eqnarray}
The second term vanishes due to being an odd function. Thus,
\begin{eqnarray}
%\begin{split}
    V(r,z,\phi)&=&-8G\mu R\cos{\phi}\int_0^{\pi}d\theta \frac{1-2\sin^2{\theta}}{\sqrt{z^2+(r+R)^2-4rR\sin^2{\theta}}}\nonumber\\
    &=&-8G\mu R\cos{\phi}\bigg(\int_0^{\pi}d\theta \frac{1}{\sqrt{z^2+(r+R)^2-4rR\sin^2{\theta}}}\nonumber\\%\right.\\
    %&\left.
    &&-2\int_0^{\pi}d\theta \frac{\sin^2{\theta}}{\sqrt{z^2+(r+R)^2-4rR\sin^2{\theta}}}\bigg).   
%\end{split}
\end{eqnarray}
From \eqref{defv}, we obtain%can cancel out $\cos{\phi}$ in each sides to recover $W(r,z)$:
\begin{eqnarray}
%\begin{split}
    W(r,z)&=&-8G\mu R\bigg(\int_0^{\pi}d\theta \frac{1}{\sqrt{z^2+(r+R)^2-4rR\sin^2{\theta}}}\nonumber\\%\right.\\
    %&\left.
    &&-2\int_0^{\pi}d\theta \frac{\sin^2{\theta}}{\sqrt{z^2+(r+R)^2-4rR\sin^2{\theta}}}\bigg).
%\end{split}
\end{eqnarray}
Taking the fact that both terms are even, and using the definitions of complete elliptic integrals, we have%$K(k)$ and $E(k)$ we have
%Defining 
%\begin{equation}
%    k\equiv\frac{4rR}{z^2+(r+R)^2},
%\end{equation}
%we get
%\begin{equation}
%    W(r,z)=-\frac{8G\mu R}{\sqrt{z^2+(r+R)^2}}\bigg(\int_0^{\pi}d\theta \frac{1}{\sqrt{1-k\sin^2{\theta}}}-2\int_0^{\pi}d\theta \frac{\sin^2{\theta}}{\sqrt{1-k\sin^2{\theta}}}\bigg).
%\end{equation}
%Both integrands are even with respect to $\pi/2$, hence
%\begin{equation}
%    W(r,z)=-\frac{16G\mu R}{\sqrt{z^2+(r+R)^2}}\left(\int_0^{\pi/2}d\theta \frac{1}{\sqrt{1-k\sin^2{\theta}}}-2\int_0^{\pi/2}d\theta \frac{\sin^2{\theta}}{\sqrt{1-k\sin^2{\theta}}}\right).
%\end{equation}
%Using the definitions of complete elliptic integrals $K(k)$ and $E(k)$ we have
%\begin{equation}
%\begin{split}
%    K(k)&=\int_0^{\pi/2}\frac{1}{\sqrt{1-k\sin^2{\theta}}}d\theta\\
%    E(k)&=\int_0^{\pi/2}\sqrt{1-k\sin^2{\theta}}d\theta,
%\end{split}
%\end{equation}
%we have
%\begin{eqnarray}
%\begin{split}
     %K(k)-E(k)&=&\int_0^{\pi/2}d\theta \frac{k\sin^2{\theta}}{\sqrt{1-k\sin^2{\theta}}}\nonumber\\
%\frac{K(k)-E(k)}{k}=\int_0^{\pi/2}d\theta \frac{\sin^2{\theta}}{\sqrt{1-k\sin^2{\theta}}},
%\end{split}
%\end{eqnarray}
%and therefore
%\begin{eqnarray}
%    \begin{split}
%        W(r,z)&=&-\frac{16G\mu R}{\sqrt{z^2+(r+R)^2}}\left(K(k)-\frac{2}{k}\left(K(k)-E(k)\right)\right)\nonumber\\
%        &=&-\frac{16G\mu R}{\sqrt{z^2+(r+R)^2}}\left(\left(1-\frac{2}{k}\right)K(k)+\frac{2}{k}E(k)\right).
%    \end{split}
%\end{eqnarray}
%Substituting back the definition of $k$, we have
\begin{eqnarray}
\label{wfin}
%\begin{split}
    W(r,z)&=&\frac{16G\mu R}{\sqrt{z^2+(r+R)^2}}\bigg(\left(\frac{z^2+(r+R)^2}{2rR}-1\right)K\left(\frac{4rR}{z^2+(r+R)^2}\right)\nonumber\\%\right.\\
    %&\left.
    &&-\frac{z^2+(r+R)^2}{2rR}E\left(\frac{4rR}{z^2+(r+R)^2}\right)\bigg).    
%\end{split}
\end{eqnarray}
%and thus we have the metric solution of $A$ expressed in complete elliptic integral
Thus, the metric function $A(r,z)$ in terms of complete elliptic integral as well as the hypergeometric function reads
\begin{eqnarray}
\label{AfinalanalyticAp}
%    \begin{split}
    A(r,z)&=&\frac{16G\mu R}{r\sqrt{z^2+(r+R)^2}}\bigg(\left(\frac{z^2+(r+R)^2}{2rR}-1\right)K\left(\frac{4rR}{z^2+(r+R)^2}\right)\nonumber\\%\right.\\
    %&\left.
    &&-\frac{z^2+(r+R)^2}{2rR}E\left(\frac{4rR}{z^2+(r+R)^2}\right)\bigg)\nonumber\\    
%\end{split}
%\end{eqnarray}
%or in the form of hypergeometric function
%\begin{eqnarray}
%    \begin{split}
&=&\frac{8\pi G\mu R}{r\sqrt{z^2+(r+R)^2}}\bigg(\left(\frac{z^2+(r+R)^2}{2rR}-1\right){}_2F_1\left(\frac{1}{2},\frac{1}{2};1;\frac{4rR}{z^2+(r+R)^2}\right)\nonumber\\%\right.\\
   % &\left.
    &&-\frac{z^2+(r+R)^2}{2rR}{}_2F_1\left(\frac{1}{2},-\frac{1}{2};1;\frac{4rR}{z^2+(r+R)^2}\right)\bigg).    %\end{split}
\end{eqnarray}

\end{appendix}

\bibliographystyle{agsm}

\end{document}